\documentclass[aip,rsi,reprint,graphicx]{revtex4-1} 

\draft 

\usepackage{amsmath}    
\usepackage{verbatim}   
\usepackage{color}      
\usepackage{subfigure}  
\usepackage{placeins}	
\usepackage[pdftex]{graphicx} 
\usepackage{epstopdf}
\usepackage{hyperref}
\usepackage{textcomp}
\usepackage[english]{babel}
\usepackage{gensymb}
\usepackage{ulem}
\usepackage{xcolor}
\usepackage{units}
\usepackage{url}
\hypersetup{
	colorlinks=true,
	linkcolor=blue,
	citecolor=blue,
	urlcolor=blue,
	breaklinks=true}
 
\begin{document}
	\title{Sub-kelvin temperature management in ion traps for optical clocks} 
	
	\author{T. Nordmann} 
	\affiliation{Physikalisch-Technische Bundesanstalt (PTB), Bundesallee 100, 38116 Braunschweig, Germany}
	\author{A. Didier} 
	\affiliation{Physikalisch-Technische Bundesanstalt (PTB), Bundesallee 100, 38116 Braunschweig, Germany}	
	\author{M. Dole{\v z}al} 
	\affiliation{Czech Metrology Institute (CMI), Okru{\v z}n{\' i} 31, 638 00 Brno, Czech Republic}	
	\author{P. Balling} 
	\affiliation{Czech Metrology Institute (CMI), Okru{\v z}n{\' i} 31, 638 00 Brno, Czech Republic}	
	\author{T. Burgermeister} 
	\affiliation{Physikalisch-Technische Bundesanstalt (PTB), Bundesallee 100, 38116 Braunschweig, Germany}
	\author{T. E. Mehlst{\"a}ubler}
	\email[]{Tanja.Mehlstaeubler@ptb.de}
	\affiliation{Physikalisch-Technische Bundesanstalt (PTB), Bundesallee 100, 38116 Braunschweig, Germany}		
	\affiliation{Institut f{\"u}r Quantenoptik, Leibniz Universit{\"a}t Hannover, Welfengarten 1, 30167 Hannover, Germany}		
	
	
	\date{\today}
	
	\begin{abstract}
		The uncertainty of the ac Stark shift due to thermal radiation represents a major contribution to the systematic uncertainty budget of state-of-the-art optical atomic clocks. In the case of optical clocks based on trapped ions, the thermal behavior of the rf-driven ion trap must be precisely known. This determination is even more difficult when scalable linear ion traps are used. Such traps enable a more advanced control of multiple ions and have become a platform for new applications in quantum metrology, simulation and computation. Nevertheless, their complex structure makes it more difficult to precisely determine its temperature in operation and thus the related systematic uncertainty. We present here scalable linear ion traps for optical clocks, which exhibit very low temperature rise under operation. We use a finite-element model refined with experimental measurements to determine the thermal distribution in the ion trap and the temperature at the position of the ions. The trap temperature is investigated at different rf-drive frequencies and amplitudes with an infrared camera and integrated temperature sensors. We show that for typical trapping parameters for $\mathrm{In}^{+}$, $\mathrm{Al}^{+}$, $\mathrm{Lu}^{+}$, $\mathrm{Ca}^{+}$, $\mathrm{Sr}^{+}$ or $\mathrm{Yb}^{+}$ ions, the temperature rise at the position of the ions resulting from rf heating of the trap stays below 700\,mK and can be controlled with an uncertainty on the order of a few 100\,mK maximum. The corresponding uncertainty of the trap-related blackbody radiation shift is in the low 10$^\mathrm{-19}$ and even 10$^\mathrm{-20}$ regime for $^{171}$Yb$^{+}$(E3) and $^{115}$In$^{+}$, respectively.
	\end{abstract}
	
	
	\maketitle 
	

	\section{Introduction}
	\label{introduction}  
	One of the major contributions to the systematic frequency uncertainty of all optical clocks is the uncertainty of the ac Stark shift induced by thermal blackbody radiation (BBR). For clocks based on neutral atoms trapped in optical lattice potentials, this requires a precise knowledge of the temperature of the vacuum chamber and environment.
	For the environment a control at the level of 5\,mK uncertainty has been demonstrated~\cite{Ludlow_2015}. 
	In the case of optical clocks based on trapped ions, the nearest surfaces to the ions are the trap electrodes at distances of a few 100\,\textmu m. rf-driven ion traps operate at several 100\,V to several kV rf amplitudes and several amperes of current flowing through the electrodes. Heating sources, dielectric losses and temperature gradients of the ion traps must be known very precisely to determine the temperature at the position of the ion with a low uncertainty. The best temperature uncertainty at the location of the ions achieved so far \cite{Nisbet-Jones_2016} is 140\,mK in a end-cap trap. Today's best optical clocks based on trapped atomic ions have uncertainties accounted to the BBR shift of 4.2~$\times$~10$^\mathrm{-19}$ and 1.7~$\times$~10$^\mathrm{-18}$ with temperature uncertainties at the location of the ion of 2.7\,K and 1.1\,K, respectively.\cite{Brewer_2019, Sanner_2019} For transitions in ions with a higher static differential polarizability such as $^{88}$Sr$^{+}$ and $^{40}$Ca$^{+}$, BBR shift uncertainties of 1.1~$\times$~10$^\mathrm{-17}$ and 1.3~$\times$~10$^\mathrm{-17}$ are reported respectively, which are dominated by the temperature uncertainty at the position of the ion.\cite{Dube_2016, Dube_2013, Cao_2017} More and more optical clocks are using composite systems for sympathetic cooling or quantum logic readout. \cite{Brewer_2019, Wan_2015, Wolf_2016, Micke_2020, Leopold_2019, Chao_2019, Ohtsubo_2019, Kozlov_2018, Ma_2020} In addition, there is an increasing interest in multi-ion clock operation to increase the signal-to-noise ratio compared to single-ion clocks.\cite{Herschbach_2012, Barrett_2015, Schulte_2016, Aharon_2019, Kaewuam_2020, Champenois_2010} For this new generation of clocks linear ion traps are needed, which are also an ideal platform for quantum computation and quantum simulation. These more complex and scalable ion traps allow multi-register operation, so that ion loading, interrogation and storage can take place simultaneously. Their complex structure usually leads to temperature gradients, which greatly complicates determining the temperature at the position of the ions precisely.\\
	\noindent In this article we present a detailed experimental and numerical analysis of the thermal behavior of chip-based linear ion traps. The achieved trap-related temperature uncertainty at the position of the ions is below 80\,mK for typical trapping parameters of mixed In$^+$/Yb$^+$ ion crystals. This uncertainty is the smallest to-date in the ion trapping community and will improve the frequency uncertainty of modern optical ion clocks.\\
	\begin{table*}[t!]
		\caption{Properties of some dielectric materials at room temperature that are used or could be used for rf ion trap construction. The values are given for different frequencies owing to a lack of data.}
		\label{tab:material_properties}
		\begin{small}
			\begin{center}
				\begin{ruledtabular}
					\begin{tabular}{lllllll} 
						& Fused silica\footnote{Valley Design \& HEBO Spezialglass} & Sapphire\footnote{Valley Design, C-Plane sapphire, values in perpendicular direction to the C plane} & Rogers & Alumina & AlN\footnote{Valley Design} & CVD-
						\tabularnewline
						&  & & (4350B)\footnote{Rogers Corporation} & ceramics (Al$_2$O$_3$)\footnote{Kyocera} &  & Diamond\footnotemark[6]\footnotetext[6]{Diamond Materials, \cite{Ibarra_1997}}
						\tabularnewline
						\toprule
						Loss tangent (x10$^{-4}$) & $<$4 (1\,MHz) & 1 (1\,MHz) & 31 (2.5\,GHz) & 1 (1\,MHz) & 4 & 10 (15\,MHz)
						\tabularnewline
						Dielectric constant & 3.8 & 9.4 & 3.5 & 9.9 & 8.9 & 5.7
						\tabularnewline
						Thermal conductivity & 1.4 & 23 & 0.7 & 32 & 175 & $>$1800
						\tabularnewline
						(W m$^{-1}$ K$^{-1}$) &   &   &   &  &  & 
						\tabularnewline
						Coefficient of thermal & 0.52 & 7 & 10-32 & 7.2 & 4.6 & 1.0
						\tabularnewline
						expansion (10$^{-6}$ K$^{-1}$) &  &  & & &  & 
						\tabularnewline
					\end{tabular}
				\end{ruledtabular}
			\end{center}
		\end{small}
	\end{table*}
	\noindent The paper is organized as follows: in section~\ref{design+model} we describe the design and numerical modeling of the ion trap and detail input parameters. The model is used to analyze the trap thermal behavior and its heat distribution, thus leading to the precise determination of the temperature at the position of the ions. The results of temperature measurements of three new traps, labeled A, B and C are presented in section~\ref{temperature measurements} and are in good agreement with the model. Using an IR camera and Pt100 sensors directly integrated into the trap structure, we analyze the scaling of the rf-induced heating of the traps with frequency and voltage amplitude. 

	\noindent In section~\ref{possible uncertainty budget} we discuss uncertainty contributions to the temperature of the trap and at the location of the ions. An estimate on the resulting trap-related uncertainty of the BBR shift for typical ion clock species is given in section~\ref{BBR_shift_ion_species}. We conclude in section~\ref{conclusion} with a summary about the achieved temperature management in the new traps.
	\section{Design and numerical model}
	\label{design+model}
	In a previous study, thermal analysis based on finite element method (FEM) models and thermal imaging with an infrared (IR) camera of various types of 3D and 2D traps have been performed\cite{Dolezal_2015, Nisbet-Jones_2016}. This study led to design recommendations for the development of ion traps with reduced thermal heating. Suitable materials with low dielectric rf losses should be used for the dielectrics. The dimensions of the conductors and the electrical capacity of the trap should be optimized to reduce Joule heating. To ensure an efficient heat removal, materials with high thermal conductivities are preferable and the thermal contacts should be optimized to offer a good thermal heat sinking. The emissivities of the parts heating during the trap operation should be kept as small as possible to avoid fluctuations of the thermal radiation visible to the ions.\\
	
	\noindent Here we focus on chip-based linear ion traps and give an extended description of the numerical model for linear chip ion traps which is in good agreement with the experimental measurements in section~\ref{temperature measurements}.	Table~\ref{tab:overview_traps_ABC} gives an overview about the three new traps A, B and C and the performed measurements. The traps B and C are currently in use to trap ions and to perform high-precision spectroscopy\cite{Keller_2019_PRA}. Trap C was tested in a chamber with a deep-IR viewport enabling IR camera measurements.
	\subsection{Selection of materials}
	\label{choice of materials}
	In the following we discuss the advantages and drawbacks of different materials that are typically used for trap fabrication. Several dielectric materials have properties fitting the previous recommendations. A material selection together with rf-related properties is listed in table~\ref{tab:material_properties}. The design of the linear ion trap, presented here, is adapted from a previous publication\cite{Herschbach_2012}. The first versions of the trap have been realized with a glass-reinforced hydrocarbon and ceramic laminate with low electrical losses (Rogers RO4350B\texttrademark), which is broadly used in industry for electrical circuitry.\\
	
	\noindent \textit{Rogers (4350B)}\\
	Custom printed circuit boards (PCB) made from Rogers 4350B can be ordered at relatively low costs. Standard PCB fabrication techniques can be used, thus affording a great flexibility in the design and the fabrication. The suitability of this material, based on the previously published design \cite{Herschbach_2012}, has already been successfully demonstrated\cite{Pyka_2014, Keller_2015}. A similar ion trap is currently installed in a setup for an Al$^+$ quantum logic clock\cite{Hannig_2019} with BBR uncertainty at the level of $4\times10^{-19}$. Drawbacks of this material are the low thermal conductivity, a lower stiffness and less precise machining compared to ceramics or crystalline materials. \\
			\begin{table*} [t!]
		\caption{Overview of the three chip traps studied in this work.
		}
		\label{tab:overview_traps_ABC}
		\begin{small}
			\begin{center}
				\begin{ruledtabular}
					\begin{tabular}{llll} 
						Trap& Setup &Characterization & Gold thickness 
						\tabularnewline
						\toprule
						A & CMI Prague & IR-measurements (sec.~\ref{exp parameters for model} and sec.~\ref{IR measurements}) + rf heating (sec.~\ref{Pt100s measurements}) & chips 1 and 4: 4.5\,\textmu m 
						\tabularnewline
						&  & & chips 2 and 3: 2.5-3.0\,\textmu m 
						\tabularnewline
						B & PTB~(1) & Calibration of integrated Pt100 sensors (sec.~\ref{calibration Pt100s} and sec.~\ref{appendix_calibration}) & all chips 4.5\,\textmu m 
						\tabularnewline
						&  & + rf-heating\cite{Keller_2019_PRA} (sec.~\ref{Pt100s measurements}) &  
						\tabularnewline
						C & PTB~(2) & Calibration of integrated Pt100 sensors (sec.~\ref{calibration Pt100s} and sec.~\ref{appendix_calibration})& all chips 4.5\,\textmu m
						\tabularnewline
						&  & + rf-heating (sec.~\ref{Pt100s measurements})	& 
						\tabularnewline			
					\end{tabular}
				\end{ruledtabular}
			\end{center}
		\end{small}
	\end{table*}

	\begin{figure*}[htbp!]
	\includegraphics[width=15.0cm]{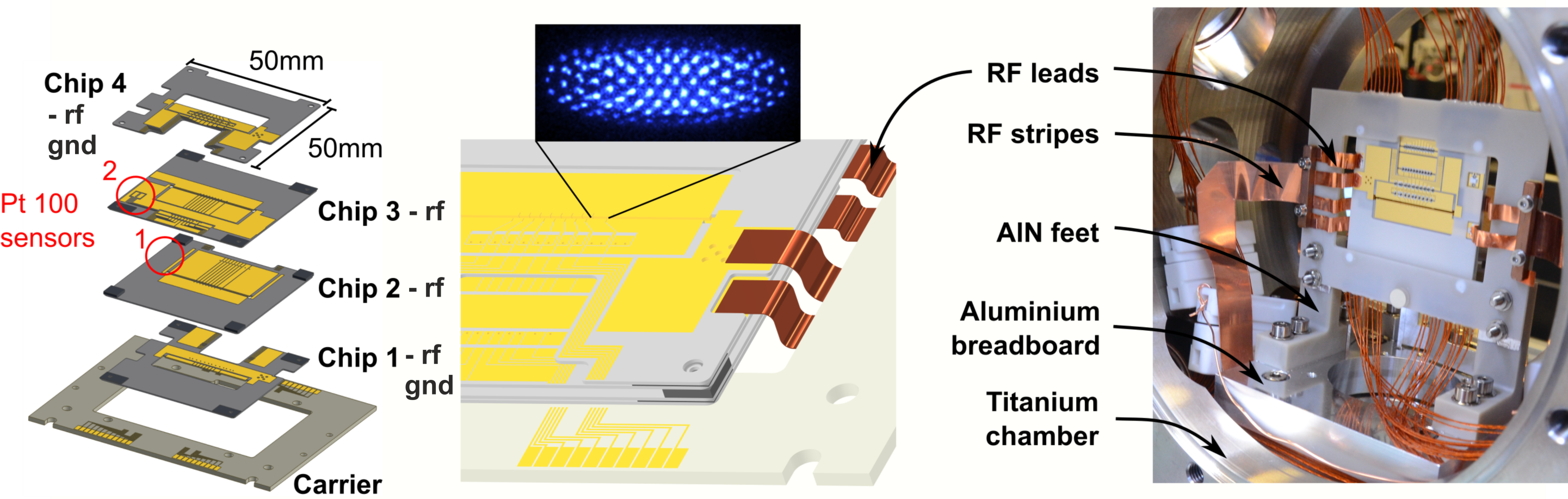}
	\caption{\textit{Left:} exploded view of the trap stack consisting of four gold-coated AlN wafers and a thick AlN support board. The rf voltage to confine the ions is applied to chips~2 and 3. The temperature of the trap is monitored with two calibrated Pt100 sensors (red circles). \textit{Middle:} 3D view of the assembled trap. \textit{Right:} trap on its carrier installed on AlN feet in a titanium vacuum chamber. A thin piece of indium foil (100\,\textmu m) is placed between all contact areas where good heat transport is critical.}
	\label{fig:trap_assembly}
	\end{figure*}
	\noindent \textit{Alumina ceramics (Al$_2$O$_3$)}\\
	An alternative and commonly used, material is aluminium oxide (Al$_2$O$_3$)\cite{Schulz_2008, Hensinger_2006}, which offers good electrical performances and thermal conductivity. It can be machined using laser ablation. A major drawback is that the ceramic material is of granular structure, which makes it more difficult to machine and displays lower homogeneity inside the material compared to crystalline materials.\\
	\noindent \textit{CVD-diamond}\\
	CVD-diamond (chemical vapor deposition) is increasingly used in industry and was recently used to fabricate ion traps\cite{Brewer_2019}. It displays very good electrical properties and an excellent thermal conductivity. It can be structured with a much higher precision than ceramics and is more homogeneous in terms of temperature gradients. It is, however, difficult and time consuming to grow. The micro-machining of graphite-free structures with high precision needs to be done with complex techniques involving fs-lasers. The crystalline material allows for extremely precise structuring and high homogeneity, however industry-grade fs-lasers became available on the market only recently. Laser processing of diamond is still under development. A high-precision diamond trap with integrated optics is currently under development at PTB together with German industry partners.\\
	\\
	\textit{Fused silica and sapphire}\\
	Another very recent development is the fabrication of 3D and 2D structures using laser-assisted chemical etching \cite{He_2014} of fused silica and sapphire. Structures with 1\,\textmu m precision in fused silica have already been demonstrated\cite{Ragg_2019}. The low electrical losses of those materials, combined with this new flexible and precise fabrication technique, make them an interesting choice for a new generation of ion traps with complex geometries. Sapphire has a better thermal conductivity than fused silica and its electrical properties are ideal for use in a cryogenic environment, but is more difficult to machine with lasers, owing to its birefringence. The first tests of  such traps are under investigation\cite{Ragg_2019, Yoshimura_2015}.\\
	
	\noindent \textit{Aluminium nitride (AlN)}\\
	The traps evaluated in this article are made of aluminium nitride (AlN), a ceramic with a high thermal conductivity and low electrical losses. This material is readily accessible and offers a low surface roughness, which leads to good thermal contacts and a low emissivity after being gold coated. A precise fabrication of the material for linear traps can be realized via ns-laser ablation. The suitability of this material for high-end traps has been demonstrated and AlN ion traps have successfully been used in the field of high-precision spectroscopy \cite{Keller_2019_PRA, Keller_Appl_2019}.\\
	\\
	All materials feature a good vacuum compatibility and have successfully been used for ion trapping. Most of the mentioned materials are classified as intrinsically ultra-high vacuum compatible in literature \cite{ligo} but without precise specification of their outgassing rate. To measure the material's outgassing rates reliably, a standardized setup is required that can resolve outgassing rates in the 10$^{-14}$\,mbar\,l\,cm$^{-2}$\,s$^{-1}$ regime and below\cite{vacom}. In the future such measurements will be carried out in a dedicated setup within the Quantum Technology Competence Center (QTZ) at PTB \cite{qtz}. The outgassing rate of a material is connected to its porosity and surface roughness. It can be expected that crystalline materials like diamond and sapphire are ideal substrates for ion trapping. Also, our experimental setups operating AlN traps exhibit excellent low collisions rates of 1 collision per 600\,s per ion. The setup with the first version of the trap based on Rogers RO4350B reached a pressure of 6~$\times$~10$^{-11}$\,mbar.\\
	\\
	\noindent All experiments presented in this article are carried out with AlN traps but the numerical model and the described  methods are valid for any kind of material. 
	\subsection{Ion trap design and passive thermal management}
	\label{passive thermal management}
	The heating behavior of a trap is influenced by numerous factors. In this section we give details on the trap design and the passive thermal management. Furthermore, the dielectric loss in the glue used for assembly is analyzed.\\
	
	\noindent The trap is composed of a stack of four AlN chips (see figure~\ref{fig:trap_assembly}). The 380\,\textmu m-thick chips are sputtered with 4\,\textmu m of gold. The outer geometry, the central slit and the conductor tracks are machined via laser cutting/structuring with a pulsed UV ns-laser. The chips are glued on AlN spacers to define distances of 1\,mm between the central chips 2 and 3, carrying the rf, and 0.127\,mm between the outer and inner chips. The distance between the ions and the electrodes is 0.7\,mm. Two calibrated Pt100 sensors are soldered to the central chips to allow the precise in-situ determination of the trap temperature (see \ref{calibration Pt100s}). The trap stack is glued on a thicker carrier made of AlN. To remove the heat generated in the trap, the carrier is screwed on AlN feet attached to a metal breadboard, which is thermally connected to the vacuum chamber. A thin sheet of indium is placed at the contact interfaces to reduce thermal resistances.\\
	
\begin{figure}[h!]
		\includegraphics[width=8.5cm]{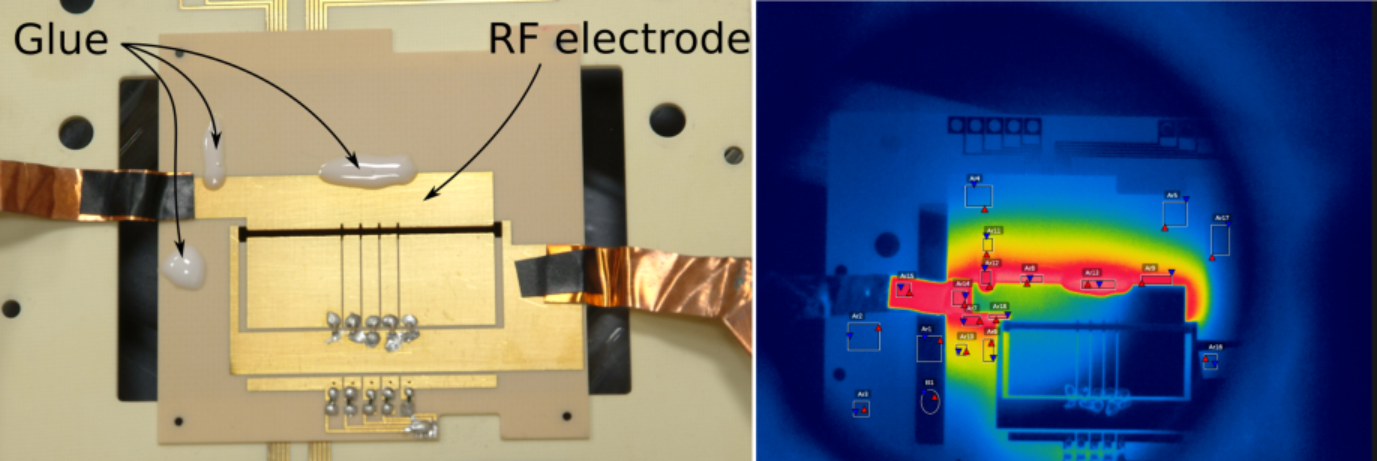}
		\caption{\textit{Left:} Picture of the chip made of Rogers where large drops of glue have been placed. \textit{Right:} IR picture showing the heat distribution when a rf voltage of 1.75\,kV amplitude at 18.6\,MHz are supplied to the chip.}
		\label{fig:test_optocast}
	\end{figure}

	\noindent \textit{Dielectric losses of the glue}\\
	We checked that the glue\footnote{Optocast 4310 Gen2} used to precisely assemble the trap does not contribute to dielectric heating. Therefore we placed large amounts of glue on a single chip made of Rogers (see figure~\ref{fig:test_optocast} left) and use a comparative measurement with an infrared (IR) camera\cite{camera} (figure~\ref{fig:test_optocast} right). A high voltage signal of 1.75\,kV amplitude at 18.6\,MHz is applied on the electrode. The temperature rise measured at the glue in direct contact with the rf electrode  is about 4.1\,K, which corresponds to about 55\,\% of the total heating. This is to be expected because the rf field is at its highest on the edges of the rf electrode and any material would increase the temperature rise if placed here. We do not see any noticeable heating of the glue placed a few millimeters away from the rf electrode.\\
	
	\noindent We conclude that the small amounts of glue used far away from the electrodes for the trap assembly do not contribute to the trap heating. Details about the experimental setup with the infrared camera can be found in section~\ref{exp parameters for model}.

	\subsection{Numerical model and experimental parameters}
	\label{exp parameters for model}
	The following section describes the numerical model of the new AlN traps (A-C) and explains which input parameters are needed. Not all emissivities of the used materials are known from literature so that their measurement and the determination of the thermal contacts inside the trap stack are described at the end of this section.\\
	\\
	\noindent \textbf{Numerical model}\\
	\begin{figure}[htbp!]
		\includegraphics[width=7.5cm]{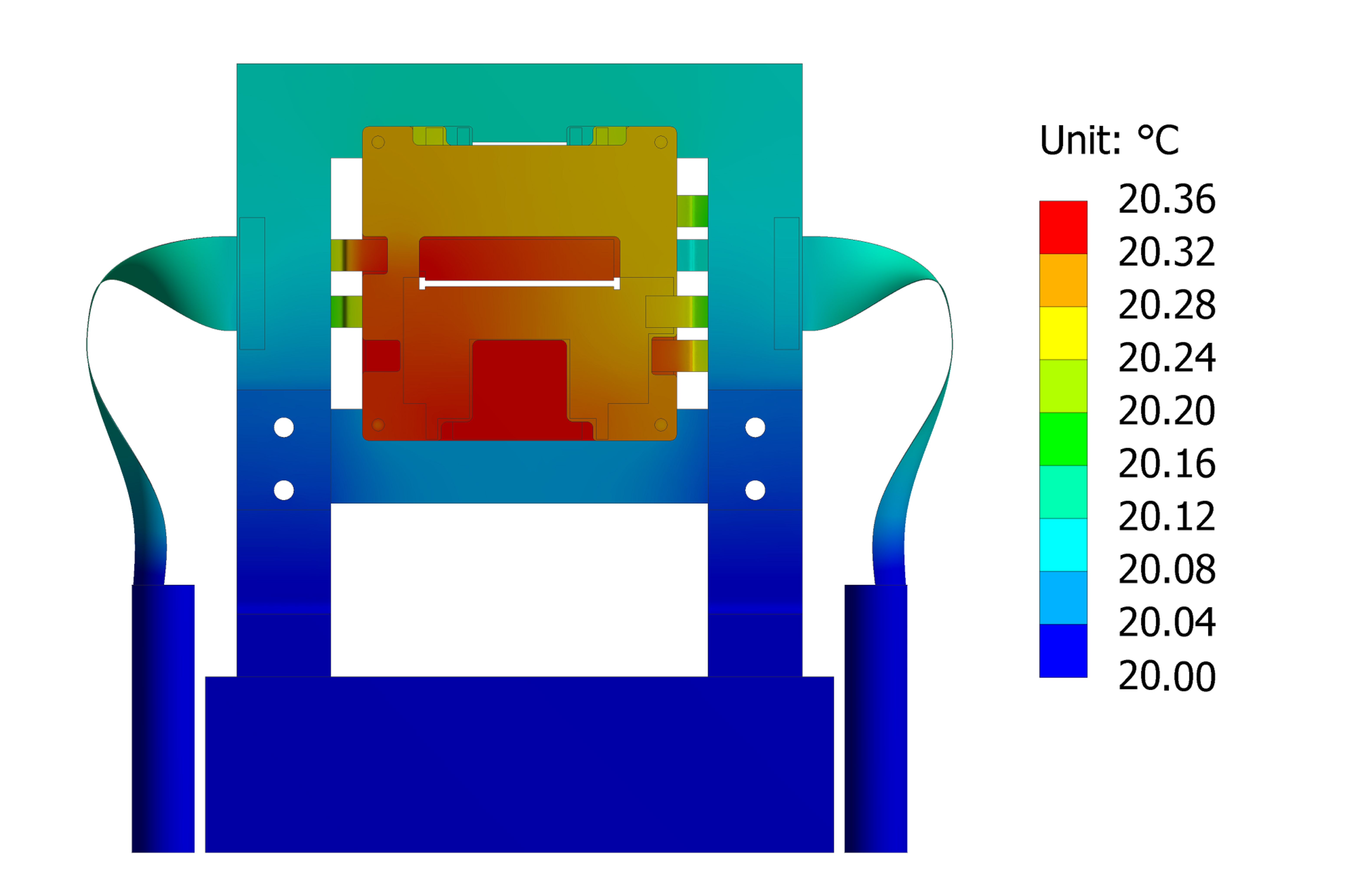}
		\caption{FEM model (ANSYS) of trap B including its AlN feet and copper strips for rf connections. The vacuum chamber is not shown for clarity.
		}
		\label{fig:FEM_trapB}
	\end{figure}

	\noindent The trap heating was modeled using the software ANSYS and follows the same method as in reference\cite{Dolezal_2015}. The geometry includes the four AlN gold-coated trap chips, the carrier along with its feet made of AlN and the copper strips used to connect the trap to the rf drive and the rf feedthrough (see figure~\ref{fig:FEM_trapB}). The feet and rf feedthrough are thermally connected to the vacuum chamber containing the trap which is set to a temperature of 20\,$\degree$C in the model. 
	
	\noindent We calculate the thermal heating arising from the rf voltage applied to the trap. The heat sources taken into account are the heat generated in the isolators as well as the Joule heating in the conductors and radiated power. The thermal heat sinking of the assembly is modeled by thermal conductivities of contacts between parts and their bulk thermal conductivities and by face to face radiation.\\

	\noindent \textbf{Input parameters for the thermal model}\\
	The required input parameters for the model are material properties such as electrical conductivities, emissivities and dielectric loss tangents of the isolators, the geometric dimensions like the thickness of the gold coating, and the thermal contacts between the parts.\\
	
	\noindent For the electrical conductivities we use literature values of the bulk materials (41\,MS/m for gold, 58\,MS/m for copper). The gold thicknesses have been measured with a profilometer and vary slightly between the three traps. For the dielectric loss tangent of AlN a value of 5~$\times$~10$^{-5}$ is used, which lies in the given range from the manufacturer and leads to agreement between the model and experiments. The emissivities of the materials surrounding the ions need to be known precisely, as they influence the magnitude of the thermal radiation to the ions.\\
	\newline	
	\textit{Emissivities}\\
	To measure the emissivities, samples of the used materials and a nearly perfect black emitter\footnote{hole in copper covered by cuprous oxide with an emissivity $>$99.5\%} are placed on a copper disk which is stabilized to temperatures between 3\,$\degree$C and 45\,$\degree$C. The setup is placed inside a vacuum chamber with a residual pressure of about 10$^\mathrm{-3}$\,mbar and IR images are taken through a ZnSe viewport. By plotting the detected signals $S$ in dependence of $T^4$ and doing a linear regression ($S$~=~ $k_\mathrm{i,ref}$ $\cdot$ $T^4_\mathrm{ref}$ + $b$) the emissivities of the samples $\epsilon_i$ can be extracted from the ratio  $\epsilon_i$ = $k_\mathrm{i}/k_\mathrm{ref}$. The additive constant is mainly related to reflected radiation. The values measured here are valid only for this set-up since the IR camera sensitivity is spectrally limited and also the ZnSe viewport transmission has a spectral dependency. The measurements agree nevertheless with published values, when available. The resulting values for the emissivities of the selected materials are listed in table~\ref{tab:emissivities}.
	\begin{table} [htbp!]
		\caption{Measured effective emissivities of different materials used for trap fabrication or calibration of the IR-measurement set-up.}
		\label{tab:emissivities}
		\begin{small}
			\begin{center}
				\begin{ruledtabular}
					\begin{tabular}{lll} 
						Material& Emissivity $\epsilon$ & $\sigma$
						\tabularnewline
						\toprule
						Shapal\cite{Dolezal_2015} & 86.7\% & 0.3\%
						\tabularnewline
						AlN polished\cite{Dolezal_2015} & 73.0\% & 0.8\%
						\tabularnewline
						gold sputtered (4\,\textmu m on AlN)\cite{Dolezal_2015} & 4.3\% & 0.8\%
						\tabularnewline
						gold electroplated (4-6\,\textmu m on AlN)\cite{Dolezal_2015} & 4.3\% & 0.8\%
						\tabularnewline
						Au/Ni/Cu leads on Rogers 4350B & 5.3\% & 0.7\%
						\tabularnewline
						Rogers 4350B & 97.8\% & 0.3\%
						\tabularnewline
						Glue\footnote{Optocast 4310 Gen2} & 90.9\% & 0.4\%
						\tabularnewline
						black tape (PVC adhesive) & 94.9\% & 0.3\%
						\tabularnewline
					\end{tabular}
				\end{ruledtabular}
			\end{center}
		\end{small}
	\end{table}
	
	\noindent For a more detailed explanation about the measurement of emissivities and obtained values for common vacuum materials the reader is referred to reference \cite{Ablewski_2020}.\\
	
\begin{figure}[h!]
		\includegraphics[width=8.5cm]{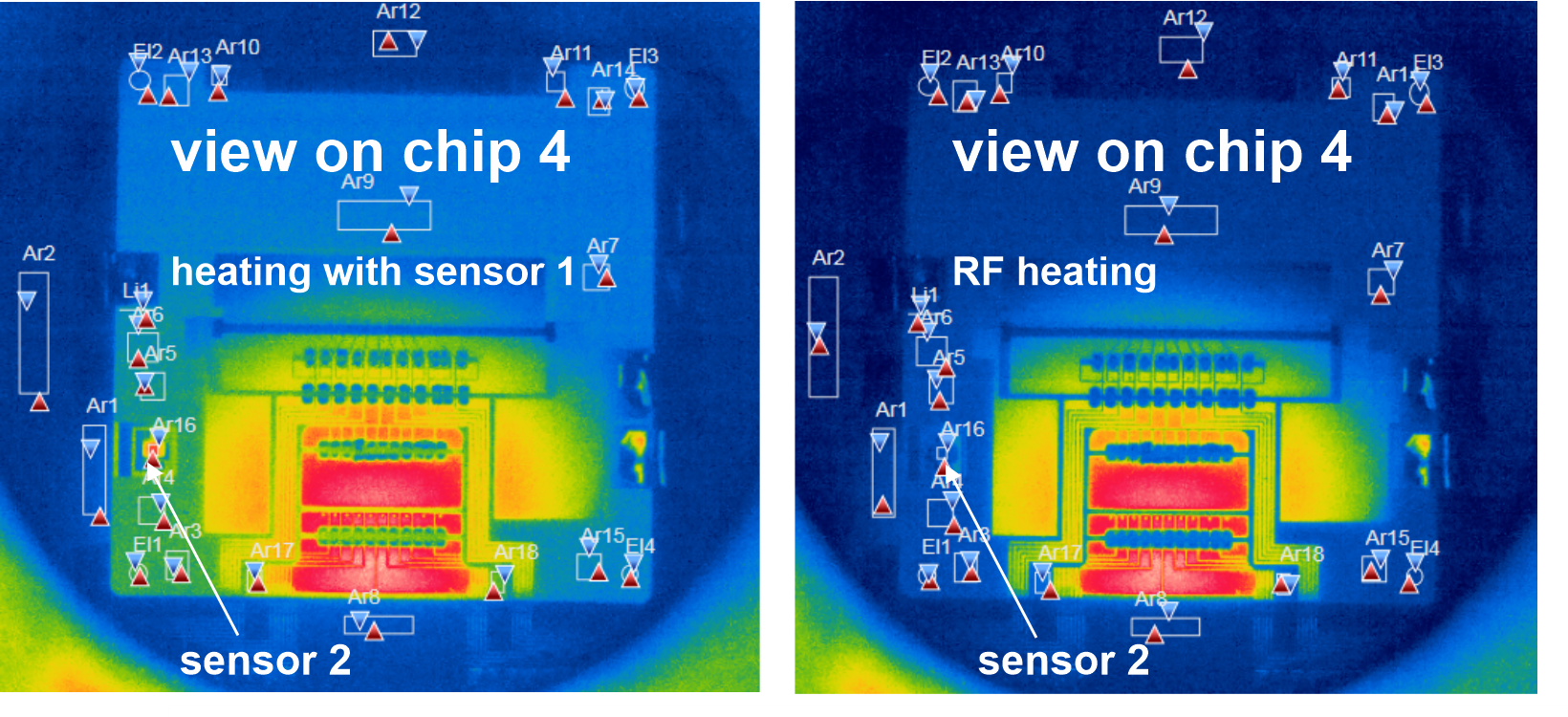}
		\caption{\textit{Left:} IR image of DC heating caused by Pt100 sensor 1 in trap~A (101\,mW), with Pt100 sensor 2 visible. \textit{Right:} Heating of trap~A through rf current with 1\,kV amplitude at 15.3\,MHz, with Pt100 sensor 2 visible. No corrections for emissivities, reflectivities nor reflected radiation are applied.}
		\label{fig:IR_images_DC_and_rf}
	\end{figure}

\begin{figure*}[htbp!]
		\includegraphics[width=16.0cm]{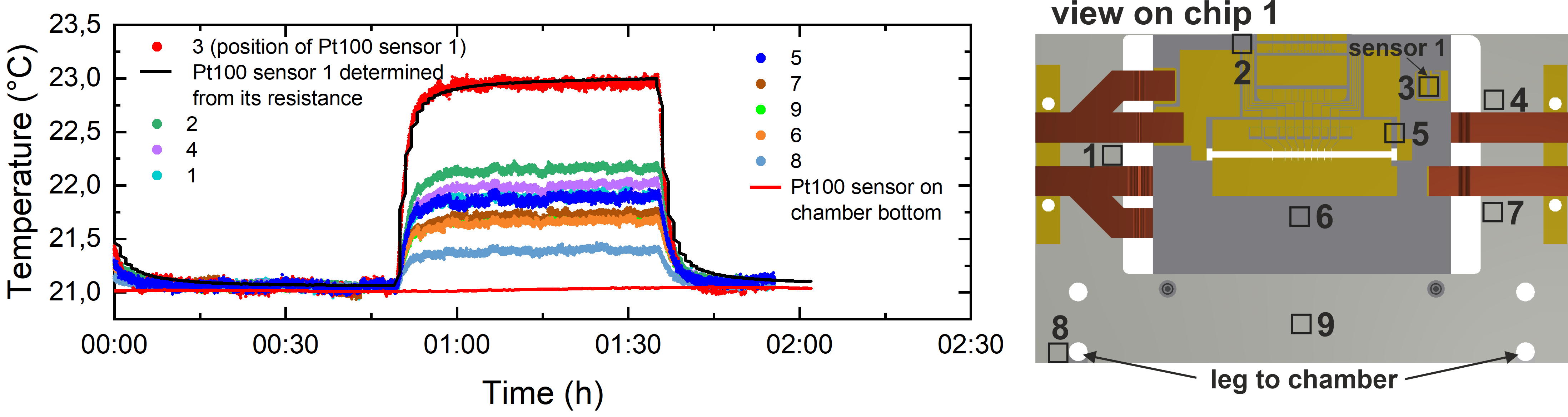}
		\caption{\textit{Left:} IR measurement for several regions of trap~A during DC heating using Pt100 sensor 2 (296\,mW) with applied corrections for emissivities and reflectivities. \textit{Right:} Scheme for the analyzed trap regions, showing the monitored positions.
		}
		\label{fig:DC_heating}
	\end{figure*}

	\noindent \textit{Thermal contacts}\\
	The precise determination of thermal contacts is complicated. It requires a known heat input and a sensor behind the thermal contact to be measured. Having two sensors placed on each central chip 2 and 3, we use one sensor as a heater by passing a DC current through it and the second as a sensor and vice versa to obtain a first estimate of the thermal resistance between the central chips. We then adjust the values of the thermal contacts between chips and spacers in the model to obtain a good agreement with the Pt100 measurements. For trap~A also parallel IR measurements are performed (see figures~\ref{fig:IR_images_DC_and_rf} and \ref{fig:DC_heating}).\\	
	
	\noindent On the left in figure~\ref{fig:IR_images_DC_and_rf} an IR image of trap~A is shown where sensor 1 (on chip 2) is used to introduce a heat load of 101\,mW and the trap is monitored with the IR camera facing chip 4 showing sensor 2 (on chip 3). The heat distribution of this DC heating with 101\,mW is comparable to the heating of the trap induced by the application of a rf voltage of 1\,kV amplitude at 15.3\,MHz as shown on the right in figure~\ref{fig:IR_images_DC_and_rf}. Corrections for emissivities, reflectivities and the loss of the vacuum window are not applied in both IR images. For details about how this is done the reader is referred to section~\ref{IR measurements}. Those uncorrected images cannot be used to determine absolute temperatures and cannot be interpreted by the apparent color due to the different material emissivities. \\
	
	\noindent Figure~\ref{fig:DC_heating} displays the evaluation of an IR measurement and the resistance measurement of sensor 2 of trap~A while applying 296\,mW of heating using sensor~1. The IR camera monitors from the direction of chip 1 thereby also showing sensor 1. The shown temperature curves belong to the regions of the trap indicated on the right of figure~\ref{fig:DC_heating}. They have been corrected with respect to emissivities, reflectivities and the loss of the vacuum window. An additional Pt100 sensor is placed on the vacuum chamber ground. The extracted temperatures of the IR camera and the Pt100 sensor 1 agree within 0.1\,K (region 3). The carrier board is colder than the chips above, showing that heat sinking could be further improved. The temperatures extracted for the multiple regions are used to determine the missing thermal conductivities within the simulation. \\

	\noindent By fitting thermal conductivities between trap chips, carrier board and legs in a range from 
	1~$\times$10$^3$\,W\,m$^{-2}$\,K$^{-1}$ and 1~$\times$10$^4$\,W\,m$^{-2}$\,K$^{-1}$ we get very good agreement between the FEM simulation and all experimental trap setups.
	
	\subsection{Thermal analysis}
	\label{thermal analysis}
	The numerical model described in the previous section (\ref{exp parameters for model}) can be used to reveal information about the distribution of heat generation among the different trap components. In this section we analyze the heat generation of this series of traps and compare it to the previous generation of traps\cite{Dolezal_2015}.\\
	
		\begin{table} [h!]
		\caption{Distribution of the heat generation among the trap components (trap~B) given by the FEM model. The values given for the absolute heating power are given for a rf voltage amplitude of 500\,V at 24.4\,MHz.}
		\label{tab:repartition_rf_heating}
		\begin{small}
			\begin{center}
				\begin{ruledtabular}
					\begin{tabular}{llll} 
						& Trap part & Power & \% of total
						\tabularnewline
						&  & (mW) & heat generated
						\tabularnewline
						\hline
						rf heating & Copper strips & 16.5 & 35.3
						\tabularnewline
						& Boards + AlN feet & 6.8 & 14.5
						\tabularnewline
						& Gold coatings & 23.2 & 49.5
						\tabularnewline
						& Others & 0.3 & 0.7
						\tabularnewline
						& \textbf{Total rf heating} & \textbf{46.8} & \textbf{100.0}
						\tabularnewline
						\hline
						Loss by & Copper strips & 0.2 & 0.5
						\tabularnewline
						radiation& Boards & 4.7 & 10.1
						\tabularnewline
						& Gold coatings & 0.2 & 0.4
						\tabularnewline
						& Others & 0.1 & 0.3
						\tabularnewline
						& \textbf{Total radiation} & \textbf{5.3} & \textbf{11.3}
						\tabularnewline
						\hline
						Conduction & Copper strips & 5.0 & 10.6
						\tabularnewline
						& Mounts & 36.5 & 78.1
						\tabularnewline
						& \textbf{Total conduction} & \textbf{41.5} & \textbf{88.7}
					\end{tabular}
				\end{ruledtabular}
			\end{center}
		\end{small}
	\end{table}

\begin{figure}[htbp!]
		\includegraphics[width=7.5cm]{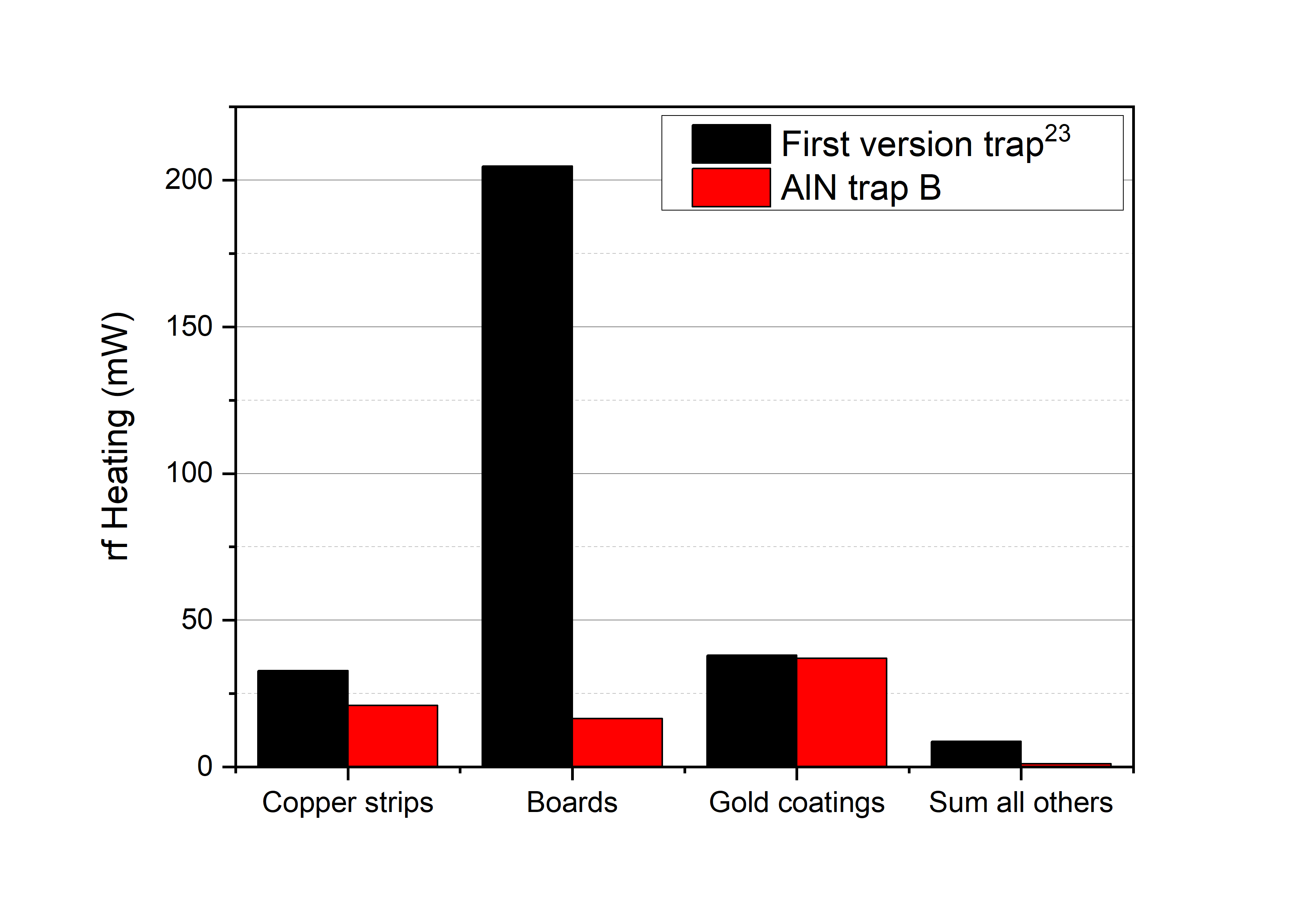}
		\caption{Distribution of the heat generation for the previous generation of the trap\cite{Dolezal_2015} (black) and this work (red) both at 1\,kV~rf amplitude at 15.4 MHz.}
		\label{fig:repartition_rf_heating}
	\end{figure}

	\noindent Table~\ref{tab:repartition_rf_heating} lists the heat generation and percentage contributions for all trap components of trap~B according to the model. The absolute values are given for an rf voltage amplitude of 500\,V and a frequency of 24.4\,MHz. 15\% of the total heat generation is attributed to the dielectrics (boards), 35\% to the copper strips supplying the trap with rf voltage and 50\% to the Joule heating in thin conductors such as the gold coating on the chips. We confirm the heat distribution given by the model with a frequency dependent analysis of trap temperature rises measured with the integrated Pt100 sensors.\\
	In figure~\ref{fig:repartition_rf_heating} we compare the distribution of the rf heating of trap~B with the first generation of the trap, which was presented in reference\cite{Dolezal_2015}. The major improvements in the rf heating lie in the better surface quality of the AlN chips and a better-controlled fabrication process. As a consequence, a much smaller amount of heat is observed in the dielectrics (boards). The dominant part of the heating process now takes place in the thin gold coatings. The other important improvements lie in the thermal contacts of the trap stack and its carrier, which have been improved per design and in the fabrication process.
	\subsection{Temperature derivation at the position of the ions}
	\label{Temperature derivation at the positions of the ions}
	The BBR at the position of the ions induces an ac Stark shift on the clock transition. To specify the shift with the smallest possible uncertainty, the temperature at the position of the ions $T_\mathrm{ions}$ needs to be known precisely. With the application of a rf voltage used to confine ions the  trap experiences a temperature rise compared to the temperature of the vacuum chamber $T_\mathrm{chamber}$. The temperature rise of the trap results in a temperature rise at the position of the ions $\Delta T_\mathrm{ions}$ because a large part of the solid angle visible to the ions is covered by the ion trap. The temperature at the position of the ions $T_\mathrm{ions}$ is therefore described by equation~\eqref{eq:T_ion}.\\
	\begin{align}
	\label{eq:T_ion}
	T_\mathrm{ions}=T_\mathrm{chamber}+\Delta T_\mathrm{ions}
	\end{align}
	The integrated Pt100 sensors allow the in-situ temperature measurement of the trap at the location of the sensors and thereby the determination of the rf-induced temperature rise at the positions of the sensors $\Delta T_\mathrm{sensor\,\,i}$ (i=1,2) with respect to the temperature of the vacuum chamber. As the model described in section~\ref{exp parameters for model} calculates the temperature rise with respect to the temperature of the vacuum chamber at any position of the trap assembly, the temperature rise at the ions $\Delta T_\mathrm{ions}$ can be compared to the temperature rise at the location of the Pt100 sensors $\Delta T_\mathrm{sensor\,\,i}$ and by that conversion factors ($\alpha$ and $\beta$) can be derived as defined in equation~\eqref{eq:T_conversion}. If the temperature of the vacuum chamber is known, this enables the in-situ determination of the temperature at the location of the ions $T_\mathrm{ions}$ and the resulting ac Stark shift by measuring the temperature rise at the positions of the sensors $\Delta T_\mathrm{sensor\,\,i}$ (i=1,2):\\
	\begin{align}
	\label{eq:T_conversion}
	\Delta T_\mathrm{ions}=\alpha \cdot \Delta T_\mathrm{sensor\,1}=\beta \cdot \Delta T_\mathrm{sensor\,2}\\
	\nonumber (\mathrm{with}~\alpha=0.43\pm0.10~\mathrm{and}~\beta=0.32\pm0.10).
	\end{align}
	A smaller temperature uncertainty can be obtained by using the mean value derived from both sensors, as given in  equation~\eqref{eq:temperature_ion}:
	\begin{align}
	\label{eq:temperature_ion}
	\Delta T_\mathrm{ions}= \frac{\alpha \cdot \Delta T_\mathrm{sensor\,1} + \beta \cdot \Delta T_\mathrm{sensor\,2}}{2}.
	\end{align}

\begin{figure}[htbp!]
		\includegraphics[width=8.0cm]{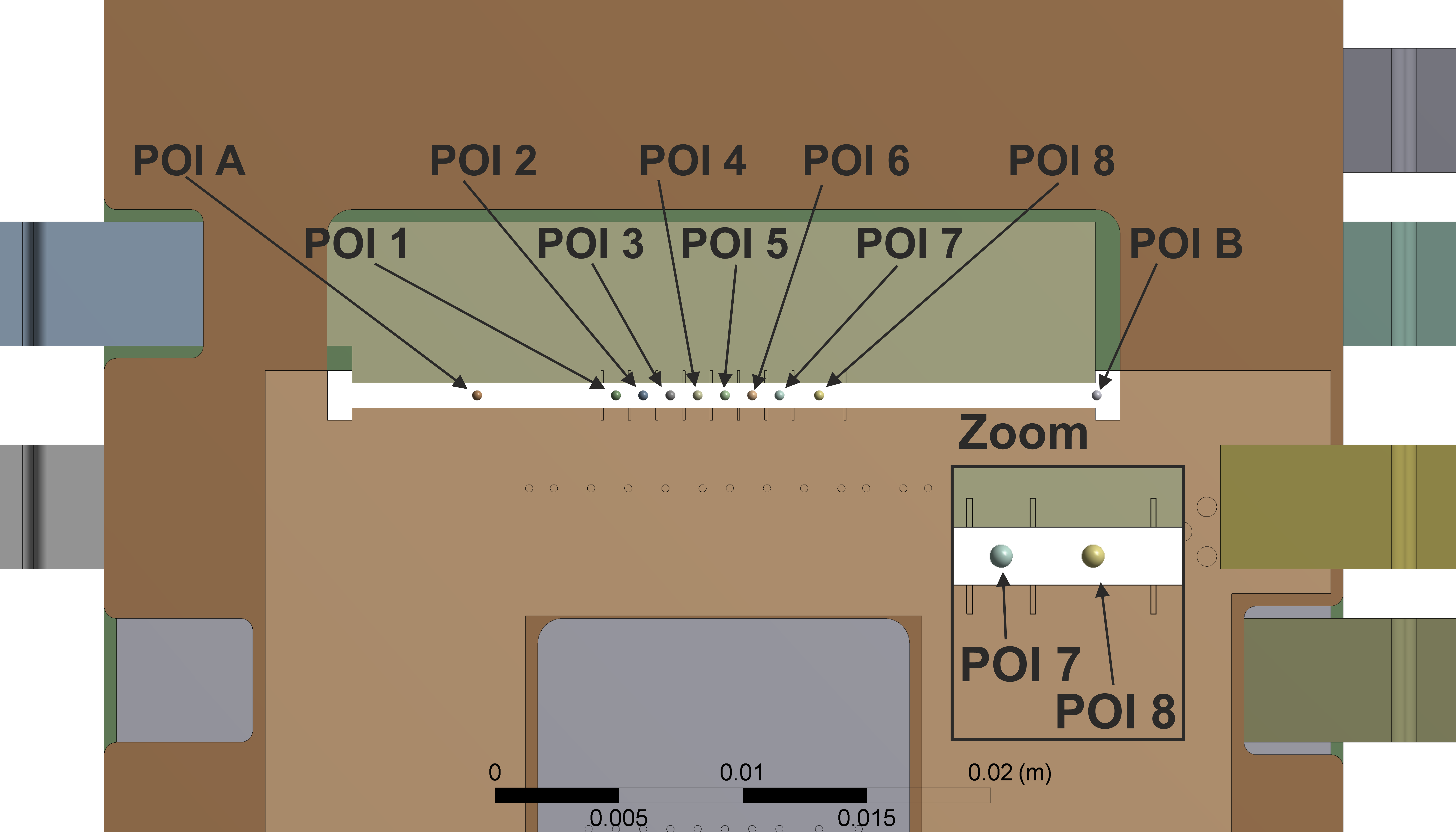}
		\caption{FEM model of the trap with several points of interest (POI) along the trap axis. The POIs are inflated to be visible.}
		\label{fig:Ansys}
	\end{figure}

	\noindent To estimate the temperature rise seen by ions stored in the trap we included small blackbody spheres (points of interest, POIs) with unity emissivity into the trap model. These spheres are positioned in the center of the trap segments, along the expected nodal line of the axial rf trapping field (see figure~\ref{fig:Ansys}). It should be noted that the POI A and POI B are not located at positions where ions can be trapped and are only used as additional sensors. 

	\noindent The uncertainties on the conversion factors in equation~\eqref{eq:temperature_ion} are deduced from 24 simulations with different values for the input parameters. The variation range of the input parameters is listed in table~\ref{tab:variation_range_parameters_FEM}. All parameters have been varied by more than 50\,\% to give a very conservative estimate. For the rf voltage uncertainty a conservative value of 10\,\% has been used. The parameter with the highest influence on the temperature at the ions is the emissivity of gold, which is coating the parts closest to the ions. A variation from 0.02 to 0.15 at a nominal value of 0.043 is taken for this input parameter to account for variations in the coating homogeneity and surface quality. The differences in temperature at the position of the ions between all 24 simulations result in an uncertainty $\sigma_{\alpha}$=$\sigma_{\beta}$=0.10 for the conversion factors $\alpha$ and $\beta$ in equation~\eqref{eq:temperature_ion}.
	
	\begin{table} [htbp!]
		\caption{Variation range of the input parameters used in the FEM model to estimate the uncertainty of the conversion factors $\alpha$ and $\beta$}.
		\label{tab:variation_range_parameters_FEM}
		\begin{small}
			\begin{center}
				\begin{ruledtabular}
					\begin{tabular}{ll} 
						Properties  & Variation
						\tabularnewline
						\toprule
						Emissivity of windows & 0.1 to 0.8
						\tabularnewline
						Emissivity gold & 0.02 to 0.15
						\tabularnewline
						Emissivity AlN &  0.5 to 1.0
						\tabularnewline
						Emissivity vacuum &  0.1 to 0.5
						\tabularnewline
						chamber & 
						\tabularnewline
						Thermal contacts (legs)&  1000 to 5000
						\tabularnewline
						(W m$^{-2}$ K$^{-1}$) &  
						 \tabularnewline
						Thermal contacts (boards) & 1000 to 6000
						\tabularnewline
						(W m$^{-2}$ K$^{-1}$) & 
						\tabularnewline
						Voltage amplitude (V) &  950 to 1050
						\tabularnewline
						Loss tangent AlN & 3~$\times$10$^{-4}$ to 9~$\times$10$^{-4}$
					\end{tabular}
				\end{ruledtabular}
			\end{center}
		\end{small}
	\end{table}
	
	\noindent The conversion factors are valid for any trapping segment of the trap within their uncertainty. The temperature rise in each segment is estimated with a different blackbody sphere in the model (see POI 1 to POI 8 in  figure~\ref{fig:Ansys}). The difference of the temperature rises between the blackbody spheres at the outer end of the trapping region (POI 1 and POI 8) and the trap center (POI 5) are below 3\,\%, thereby well below the contribution of the uncertainty of the conversion factors.\\

	\noindent In case of trap~C the temperature of the vacuum chamber is constantly monitored with seven additional Pt100 sensors, glued on the chamber, that show a maximal difference of 170\,mK under trap operation which is on the same size as the uncertainty of the sensors.
	\section{Temperature measurements}
	\label{temperature measurements}
	In this section we report about temperature measurements of all three traps. Table~\ref{tab:overview_traps_ABC} (in section~\ref{design+model}) gives an overview about the small differences between the three traps A, B and C and about the performed measurements or calibration.\\	
	
	\noindent In the first part of the section we report about the calibration of the integrated Pt100 sensors, carried out to improve the uncertainty specified by the manufacturer and to consider the circuitry on the trap. Afterwards we describe temperature measurements performed with an IR camera and the integrated Pt100 sensors to investigate temperature gradients and the scaling of the rf-induced heating of the trap with rf voltage amplitude and frequency, respectively.
	\subsection{Calibration procedure of the integrated Pt100 sensors}
	\label{calibration Pt100s}
		\begin{figure}[h!]
		\includegraphics[width=8.0cm]{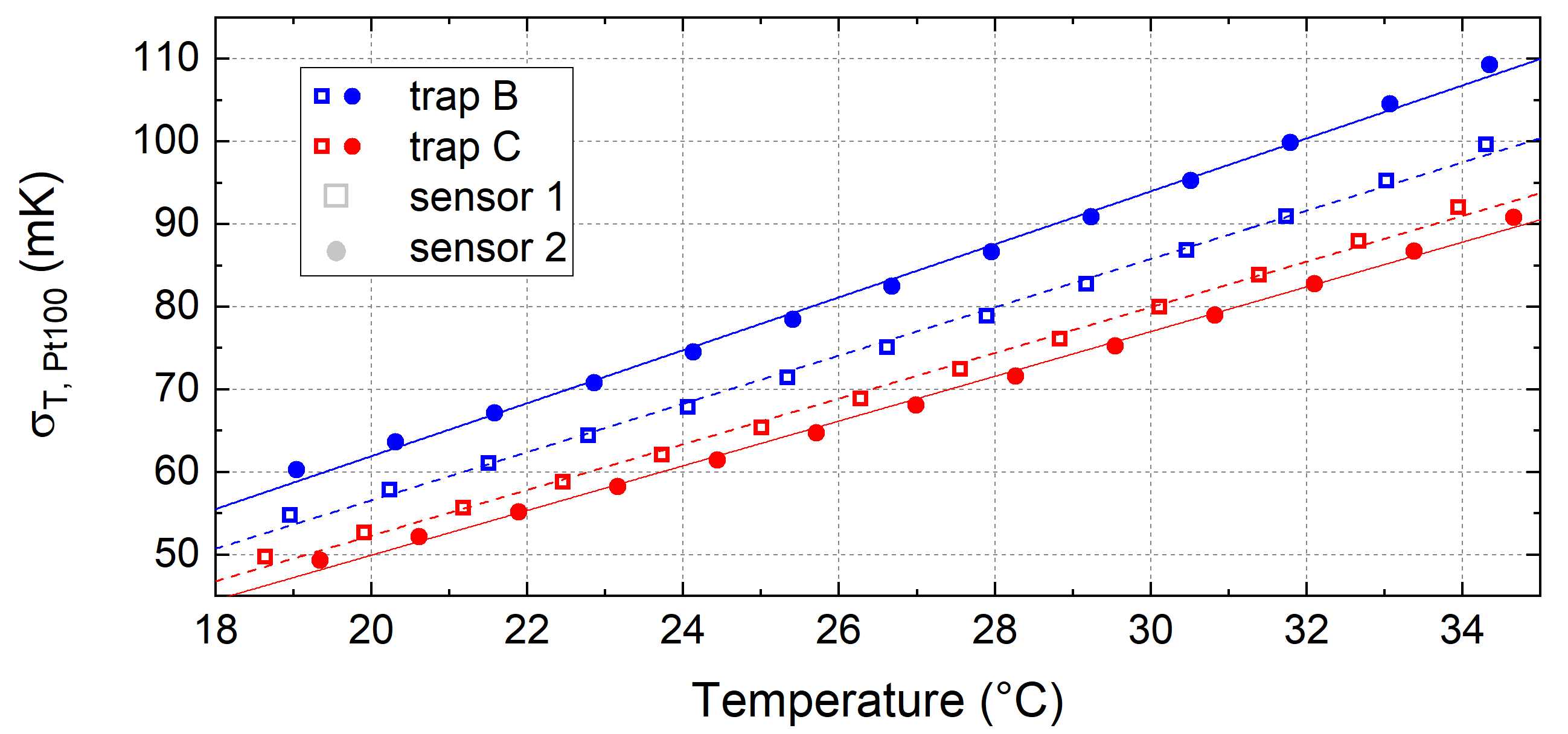}
		\caption{Absolute temperature uncertainty $\sigma_\mathrm{T, Pt100}$ for the two integrated and calibrated Pt100 sensors on the traps B (blue) and C (red). The open symbols and dashed lines correspond to sensor~1 and the closed symbols and the solid lines to sensor~2. The uncertainty is temperature-dependent and increases with about 3\,mK/$\degree$C. 
		}
		\label{fig:absolute_temperature_uncertainty}
	\end{figure}
	The two Pt100 platinum resistance temperature sensors that are integrated on the two inner rf chips (see figure~\ref{fig:trap_assembly}) can be read out via a 4-wire measurement. From this local temperature measurement the temperature at the position of the ions and thereby the resulting BBR shift can be estimated in-situ as described in equation~\eqref{eq:temperature_ion} in section~\ref{Temperature derivation at the positions of the ions}. 
	Two capacitors are placed in parallel next to the two Pt100 sensors to short all ac voltages. From the manufacturer the temperature uncertainty of the Pt100 sensors is specified with 300\,mK. For a precise temperature estimate a smaller uncertainty is essential. Therefore we calibrated the sensors integrated in trap~B and C as described in the appendix~\ref{appendix_calibration}.

	\noindent The result of the calibrations is shown in figure~\ref{fig:absolute_temperature_uncertainty} and reveals an absolute temperature uncertainty $\sigma_\mathrm{T, Pt100}$ of  below 70\,mK for both sensors in the case of trap~B and below 60\,mK for trap~C for a typical lab temperature of 22\,$\degree$C. The temperature resolution of the reading of the integrated Pt100 sensors is below 1\,mK, which allows the in-situ detection of very small temperature changes during trap operation.
	\subsection{Measurements with IR camera}
	\label{IR measurements}
	The integrated Pt100 sensors of the trap show local temperatures at the position of the sensor. To observe temperature gradients across the trap infrared measurements have been performed with trap~A. In the following the procedure and evaluation of these measurements is explained.\\
	
	\noindent Trap A is placed inside a vacuum chamber featuring a zinc selenide (ZnSe) vacuum viewport and several sensors to monitor the temperature of the chamber, the viewports and the IR camera\cite{camera}. A reference black emitter is placed at the chamber bottom next to the trap.\\ 
	
	\noindent When looking through the IR camera at an object consisting of several materials in thermal equilibrium, the apparent temperature of the different materials is different. The reasons are their different effective emissivities (depending on material and surface texture) and reflections of thermal radiation coming from other objects. For a correct evaluation, the emissivities of all materials under test must be known precisely and the reflected radiation from other objects must be subtracted from the IR image. The procedure to determine the emissivities has been described in section~\ref{exp parameters for model}.\\
	
	\noindent We use the IR camera measurements to extract the temperature rise across the whole trap induced by rf heating to analyze the maximal temperature gradient across the trap. Therefore the temperature of the chamber, trap and black emitter are monitored with the IR camera and Pt100 sensors while the trap drive is first switched off until the whole system reaches thermal equilibrium. The objects surrounding the trap are temperature stabilized to a level of 0.2\,$\degree$C so that their radiated power shining on the trap stays as constant as possible. This is later subtracted from the recordings of the camera such that the values agree with the Pt100 sensors in thermal equilibrium after also accounting for the emissivities and the viewport losses. The temperature of the bottom of the chamber recorded by a Pt100 sensor is used as a temperature reference during the measurements. The uncorrected IR reading of the black emitter at the bottom of the chamber is therefore adjusted to match this temperature, thus correcting camera drifts. After reaching thermal equilibrium the rf is switched on until the temperature is in equilibrium again. At this point the rf field is switched off again and the system is allowed to reach equilibrium one more time. This enables checking of possible residual uncorrected drifts of the camera, which are below 0.1\,$\degree$C as shown for the measurement in figure~\ref{fig:DC_heating}. With all the described steps the camera recordings are corrected for temperature drifts, reflected radiation, emissivities of the materials and for the losses attributed to the ZnSe viewport. The reflected radiation that is subtracted can be higher when the temperature is high and can lead to an overestimation of the temperature rise measured with the IR camera. This uncertainty is below the noise of the IR camera for all materials considered in this study. The uncertainty of the temperature estimation from an IR measurement increases for low emitting materials, because they are highly reflecting. To enhance the resolution and to reduce the uncertainty, thin, black tape with a thickness of 0.1\,mm with a calibrated emissivity (95\%) can be placed in that region. The achieved temperature uncertainty of the IR camera measurements is $\sigma_\mathrm{T,cam}$~=~0.2\,K +\,0.05~$\Delta T$, where $\Delta T$ is the observed temperature rise in kelvins.\\
		\begin{figure}[htbp!]
		\includegraphics[width=8.0cm]{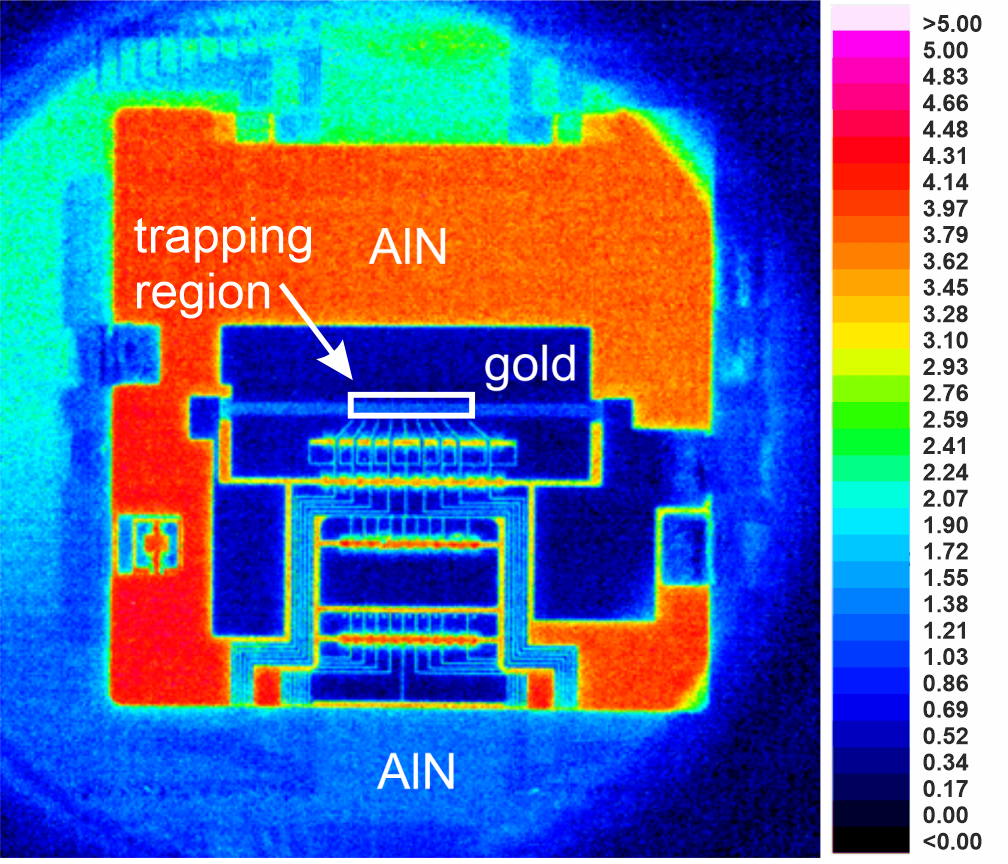}
		\caption{IR camera image of the temperature rise of trap~A induced by the application of a rf voltage at $f_\mathrm{rf}$~=~21.5\,MHz and $U_\mathrm{rf}$~=~1.5\,kV. The vacuum viewport and IR camera face the trap chip 4, with the other chips being mostly covered by it. For the shown image a cold image (rf off) has been subtracted from the hot image (rf on) and corrections for the AlN emissivity and the vacuum losses are included. This means the temperature scale is just valid for the AlN parts.
		}
		\label{fig:IR_measurement_trap_A_1.5kV}
	\end{figure}	

	\noindent Figure~\ref{fig:IR_measurement_trap_A_1.5kV} displays the temperature rise of trap~A being in thermal equilibrium at an applied rf voltage of 1.5\,kV peak at a drive frequency of $f_\mathrm{rf}$~=~21.5\,MHz compared to thermal equilibrium when no rf is supplied. We subtract the cold image, where the rf voltage is switched off for a long time, from the hot image, where the rf voltage is applied. The size of the ZnSe vacuum viewport limits the visible area of the trap such that the complete carrier board cannot be seen. The trap is placed such that chip 4 is facing the IR camera.  The shown temperature scale accounts for the emissivity of AlN and the vacuum viewport transmission. This implies that only the temperature of the AlN surfaces are shown correctly in this picture, but it can of course be adapted to the other materials. Since gold features a much smaller emissivity than AlN ($\epsilon_\mathrm{Au}$~=~0.05 and $\epsilon_\mathrm{AlN}$~=~0.78), meaning that less thermal radiation is emitted, the gold electrodes seem to be much colder as they are in reality. The comparably low emissivity of gold helps to reduce the BBR seen by the ions.
	The shown temperature rise of trap~A in figure~\ref{fig:IR_measurement_trap_A_1.5kV} is highest in the lower left corner and lowest in the upper right in the scope of that measurement and agrees with the model in figure~\ref{fig:FEM_trapB}. The difference between these two points is 0.51\,K and is the largest gradient in that measurement. The hottest point of the whole trap is on chip~3 and the total gradient is greater than 4\,K between the trap feet and the hottest point on chip~3 according to the model.
	\subsection{Dependency of trap heating on \textbf{$U_\mathrm{rf}$} and \textbf{${f}_\mathrm{rf}$}}
	\label{Pt100s measurements}
	The scaling of the trap temperature rise with rf voltage follows a quadratic behavior~\cite{Dolezal_2015} while the scaling with the frequency is more complex. The reason for that are the different heating mechanisms in conductors and insulators. In this section we derive the frequency scaling of the different heating mechanisms theoretically and verify this scaling experimentally. Furthermore we measure the voltage scaling of the rf heating and compare the three traps. \\
	\begin{figure*}[htbp!]
		\includegraphics[width=13.0cm]{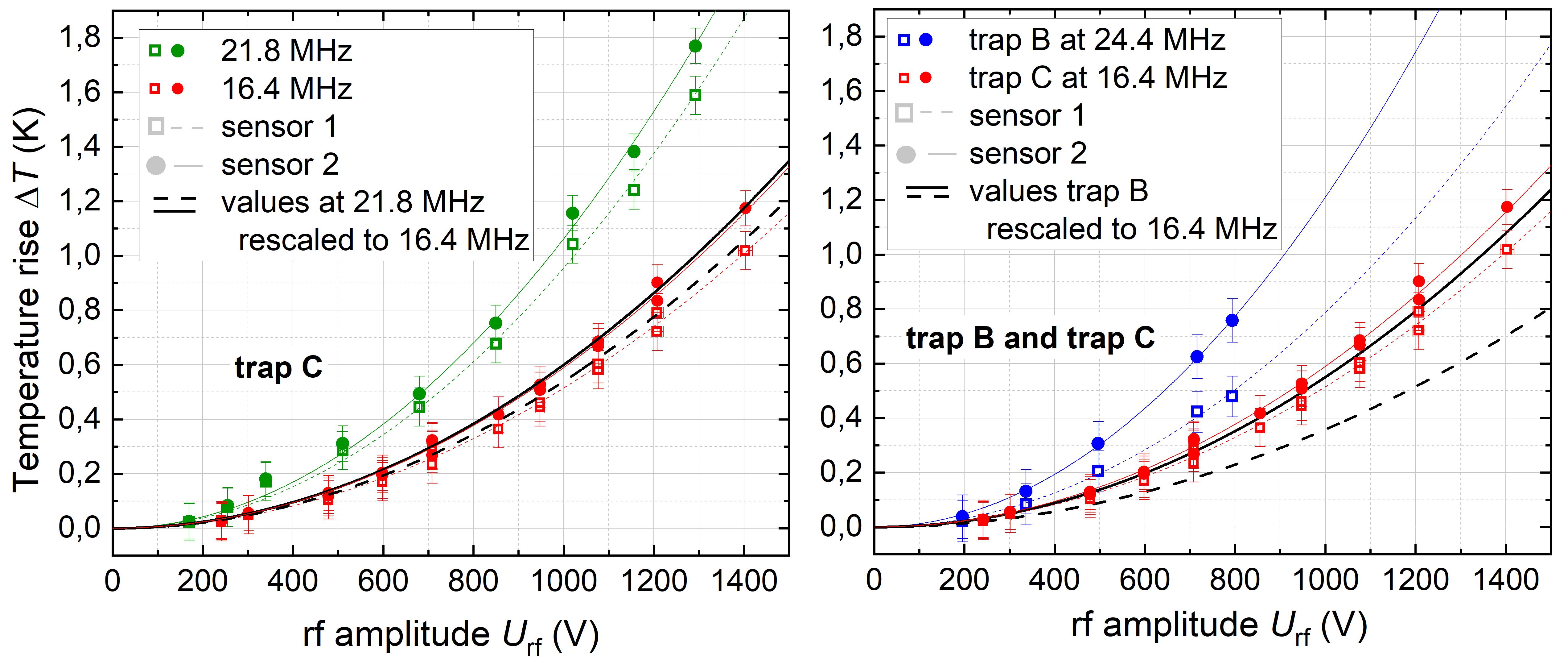}
		\caption{Voltage-dependent rf-induced temperature rise measured with the integrated Pt100 sensors. The open rectangular and closed circle symbols correspond to sensor~1 and sensor~2 respectively. The dashed and solid lines are quadratic fits to the measured values. \textit{Left:} Temperature rise vs. rf voltage in trap~C for two different rf drive frequencies: 16.4\,MHz (red) and 21.8\,MHz (green). The values at 21.8\,MHz have been rescaled to a rf frequency of 16.4 MHz with equation~\eqref{eq:rf_heating_partial} (black lines). \textit{Right:} Comparison of the  temperature rise in trap~B at 24.4\,MHz (blue) and trap~C (red), for trap~B the values have been rescaled to a rf frequency of 16.4 MHz with equation~\eqref{eq:rf_heating_partial} (black lines). The error bars are dominated by the absolute uncertainty of the integrated Pt100 sensors since the fitting errors for the warming and cooling of the trap are negligible.
		}
		\label{fig:rf_heating_voltage_dependency}
	\end{figure*}
	\\
	\textbf{Heating mechanisms and their frequency scaling}\\
	The dissipated power due to rf absorption in dielectrics is given by the relation in equation~\eqref{eq:freq_dependency_dielectrics}. Here, $E$ is the electric field induced by the drive voltage $U_\mathrm{rf}$cos($\Omega_\mathrm{rf}t$) at the angular frequency $\Omega_\mathrm{rf}$ and $\epsilon_0$ the permittivity of free space. In the case of AlN the relative permeability $\epsilon_r$ and the loss tangent tan($\delta$) are constant for the considered frequency range\cite{thorp_1990}, so that the absorbed power scales linearly with the frequency:\\
		\begin{align}
		\label{eq:freq_dependency_dielectrics}
		P_\mathrm{dielectrics}(\Omega_\mathrm{rf})=\Omega_\mathrm{rf} \cdot \epsilon_0 \cdot \epsilon_r \cdot \tan(\delta) \cdot \left|E\right|^2.
	\end{align}
  	\noindent For resistive losses, originating from conductors, the frequency dependence of the current and the resistance have to be taken in account and result in the relation given in equation~\eqref{eq:freq_dependency_res} where $\zeta(\Omega_\mathrm{rf})$~=~$\sqrt{\frac{2\rho}{\Omega_\mathrm{rf}\cdot \mu}}$is  the skin depth, \textit{d} is the thickness of the conductor, $\rho$ is the resistivity of the conductor material and $\mu$ the permeability.
	\begin{align}
	\label{eq:freq_dependency_res}
	P_\mathrm{res}(\Omega_\mathrm{rf})\propto \frac{\Omega_\mathrm{rf}^2}{\zeta(\Omega_\mathrm{rf}) \cdot  [1-\exp(-d/\zeta(\Omega_\mathrm{rf}))]}
	\end{align}
	
	\noindent For thick conductors ($d\gg\zeta(\Omega_\mathrm{rf}$)) the exponential term can be neglected and $P_\mathrm{res}\propto\Omega_\mathrm{rf}^{2.5}$. At $\Omega_\mathrm{rf}$~=~2$\pi$~$\times$~24.4\,MHz  the skin depth is 15.9\,\textmu m for gold and 13.4\,\textmu m for copper. This condition is fulfilled for the copper strips (d~=~100\,\textmu m) connecting the electrical feedthrough to the trap rf electrodes. For thin conductors like the trap electrodes of 4\,\textmu m gold, where d $\approx\zeta(\Omega_\mathrm{rf})$ the skin effect can be neglected and $P_\mathrm{res}~\propto~\Omega_\mathrm{rf}^{2}$.\\
	\noindent In summary the frequency dependence of the trap temperature rise due to rf heating can be described with equation~\eqref{eq:rf_heating_partial}. 
		\begin{align}
	\label{eq:rf_heating_partial}
	\Delta T (f_\mathrm{rf}) = \gamma_\mathrm{T} \cdot U_\mathrm{rf}^2 \cdot (\lambda_1 \cdot \frac{f_\mathrm{rf}}{f_\mathrm{ref}} + \lambda_2 \cdot  \frac{f_\mathrm{rf}^{2}}{f_\mathrm{ref}^{2}} + \lambda_3 \cdot \frac{f_\mathrm{rf}^{2.5}}{f_\mathrm{ref}^{2.5}})
	\end{align}
	\noindent $\gamma_\mathrm{T}$ describes the dependence of the temperature rise on the rf voltage amplitude $U_\mathrm{rf}$ at a reference frequency $f_\mathrm{ref}$ and can be slightly different even at the same $f_\mathrm{ref}$ for the three traps due to variations in the heat conductivity or the gold thickness originating from the manufacturing process. The coefficients $\lambda_1$, $\lambda_2$ and $\lambda_3$ are determined by the portion of the respective heating mechanism ($\lambda_1$:~dielectrics, $\lambda_2$:~thin conductors and $\lambda_3$:~thick conductors) and have to be given for the same reference rf frequency $f_\mathrm{ref}$. $f_\mathrm{ref}$ is the typical rf drive frequency used in the setups of the traps A, B and C. In principle $\gamma_\mathrm{T}$, $\lambda_1$, $\lambda_2$ and $\lambda_3$ are not frequency-dependent and could be combined and given at one reference rf frequency for all traps. For an easier comparison with the experiments we decided to give the coefficients at those setup specific reference frequencies.\\

	\noindent Table~\ref{tab:portions_heating} summarizes the coefficients $\lambda_1$, $\lambda_2$ and $\lambda_3$ derived from the heat distribution given by the model at the respective reference frequency $f_\mathrm{ref}$. The values for trap~A are slightly different because of the smaller gold thickness on the central electrodes.
	\begin{figure*}[htbp!]
		\includegraphics[width=12.5cm]{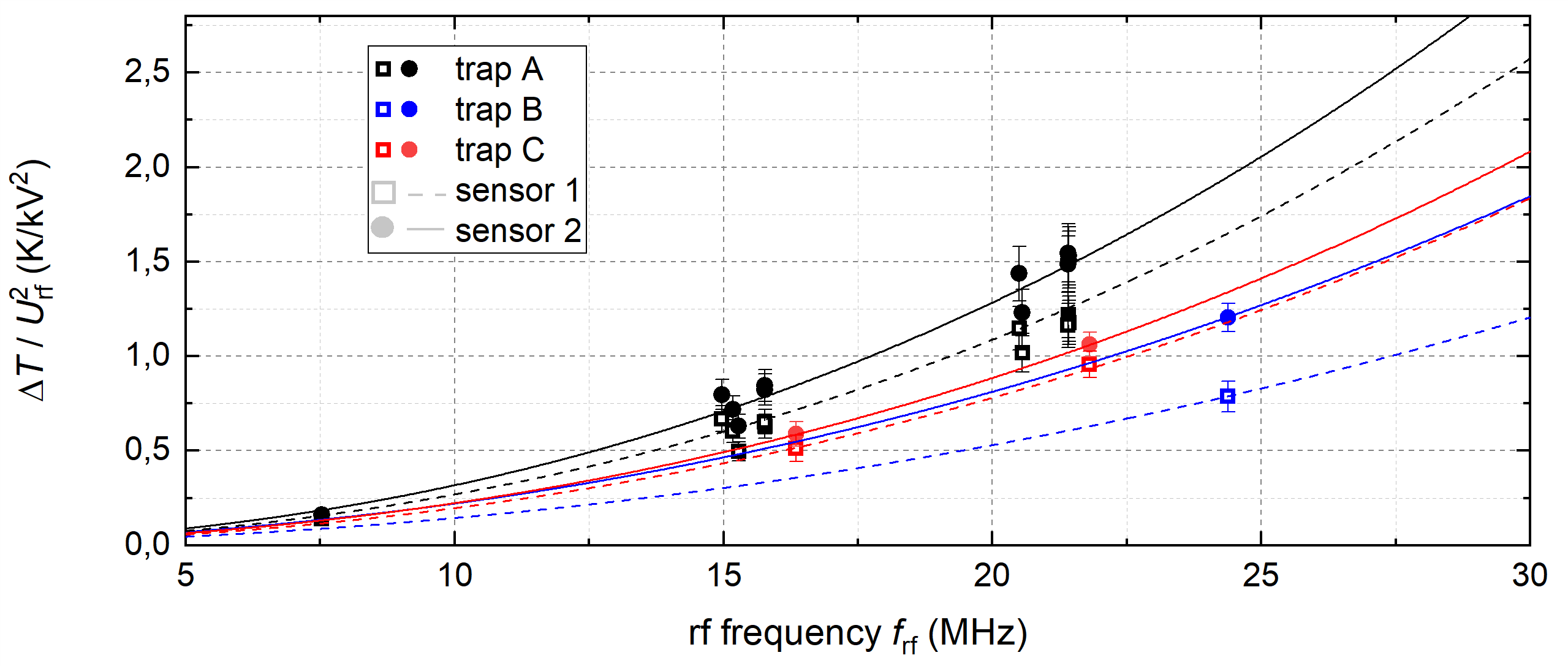}
		\caption{Frequency dependent trap temperature rise per kV$^2$ rf voltage amplitude measured with the Pt100 sensors in trap~A (black), trap~B (blue) and trap~C (red). The open symbols correspond to sensor~1 and the closed symbols to sensor~2. Trend lines according to equation~\eqref{eq:rf_heating_partial} are indicated (dashed for sensor~1 and solid for sensor~2) and show excellent agreement with the measurements.
		}
		\label{fig:rf_heating_frequency_dependency_1000V}
	\end{figure*}
	
	\begin{table} [htbp!]
		\caption{Coefficients $\lambda_1$, $\lambda_2$ and $\lambda_3$ describing the portions of the different heating mechanisms at the reference frequency $f_\mathrm{ref}$. The values are derived from the heat distribution given by the model. Trap A has different coefficients because of the thinner gold thickness.}
		\label{tab:portions_heating}
		\begin{small}
			\begin{center}
				\begin{ruledtabular}
					\begin{tabular}{lllll} 
					Trap & $f_\mathrm{ref}$ (MHz) & $\lambda_1$ & $\lambda_2$ & $\lambda_3$ 
					\tabularnewline
					\hline
					A & 15.1 & 0.14 & 0.52 & 0.34
					\tabularnewline
					B & 24.4 & 0.15 & 0.50 & 0.35
					\tabularnewline
					C & 16.4 & 0.15 & 0.50 & 0.35
					\end{tabular}
				\end{ruledtabular}
			\end{center}
		\end{small}
	\end{table}
	
	\noindent We use those coefficients for rescaling the voltage-dependent temperature rise measured at different rf frequencies, thus enabling comparisons between the three traps. \\

	\noindent \textbf{Measured $U_\mathrm{rf}$ and $f_\mathrm{rf}$ scaling}\\
	The rf-induced temperature increase of all three traps has been measured with the two integrated Pt100 sensors for several rf voltage amplitudes and frequencies (see figure~\ref{fig:rf_heating_voltage_dependency} and \ref{fig:rf_heating_frequency_dependency_1000V}). The temperature drift of the vacuum chamber over the course of the measurements has been monitored with seven additional Pt100 sensors glued on the vacuum chamber in the case of trap C and one in case of trap A and B. The respective drifts have been considered in all measurements. All three traps show extremely low heating of below 2\,K for all tested rf drive frequencies. The temperature rise on sensor 1 (positioned on chip 2, open rectangular in all figures) is always lower than on sensor 2 (positioned on chip 3, filled circle in all figures) because its location is closer to the carrier board and thereby the heat sink.\\ 

	\noindent The measurements of the $U_\mathrm{rf}$ scaling of the trap heating are shown in figure~\ref{fig:rf_heating_voltage_dependency}. To determine the temperature rise the trap drive was first switched on until the system reached thermal equilibrium and then switched off until thermal equilibrium was reached again. This warming and cooling of the trap was then fitted and the difference between the two steady-state temperatures is taken as the temperature rise $\Delta\,T$. On the left the rf voltage dependent temperature increase for trap~C is shown for two frequencies and the data have been fitted with a quadratic function: $f_\mathrm{rf}$~=~16.4\,MHz (red) and $f_\mathrm{rf}$~=~21.8\,MHz (green). The measurement for $f_\mathrm{rf}$~=~21.8\,MHz is additionally displayed after rescaling it to $f_\mathrm{rf}$~=~16.4\,MHz with equation~\eqref{eq:rf_heating_partial} (black lines) in order to verify its correctness and thereby our model. From the measurements it can be seen that the temperature rise increases with frequency which is explained by equation~\eqref{eq:freq_dependency_dielectrics} and \eqref{eq:freq_dependency_res} showing that all heating processes increase with frequency. The measurements at $f_\mathrm{rf}$~=~16.4\,MHz and $f_\mathrm{rf}$~=~21.8\,MHz overlap almost perfectly after rescaling with equation~\eqref{eq:rf_heating_partial}, implying that the partial heating processes in the trap are well understood and the model and the measurements are in agreement.\\
	
	\noindent In figure~\ref{fig:rf_heating_voltage_dependency} (right) the rf voltage-dependent temperature increase for the traps B and C is compared. Trap B is operated at $f_\mathrm{rf}$~=~24.4\,MHz (blue). To make the results comparable to trap~C the measurement of trap~B is also shown after rescaling it to $f_\mathrm{rf}$~=~16.4\,MHz with equation~\eqref{eq:rf_heating_partial} (black lines). For sensor~2 the temperature rises in traps B and C agree, while for sensor~1 they differ noticeably.  While for trap~C both temperature sensors display similar values (deviation $\approx$7\% at 1.4\,kV), the temperature sensor 1 in trap~B shows much colder temperatures ($\approx$30\% deviation at 1.4\,kV). We suspect that this difference is due to the better thermal sinking with the vacuum chamber in the set-up of trap~B and to a different geometry of the rf strips connecting the trap to the chamber rf feedthrough. FEM simulation showed qualitatively that to the latter has an influence over the spread in temperature rise between the sensor 1 and 2.
	Nevertheless the agreements in sensor 2 between trap~B and C show that the trap fabrication works reproducibly with regard to the thermal properties.\\
	
	\noindent From the quadratic fits to the measurements the scaling factor $\gamma_\mathrm{T,\,\,sensor\,\,i}$ (i=1,2 for the two sensors) giving the relation between heating and applied rf voltage at the location of the sensors 1 and 2 has been extracted for all three traps. Table~\ref{tab:rf_heating_trap_B_and_C} summarizes these values for all three traps which we measured at different rf drive frequencies ($f_\mathrm{ref}$ as explained previously).\\
	\begin{table} [htbp!]
		\caption{Scaling factors $\gamma_\mathrm{T}$ for the relation between temperature rise and rf voltage amplitude at the location of the sensors 1 and 2 for all three traps. The factors have been deduced from the measured rf voltage scaling of the temperature rise shown in figure~\ref{fig:rf_heating_voltage_dependency}.}
		\label{tab:rf_heating_trap_B_and_C}
		\begin{small}
			\begin{center}
				\begin{ruledtabular}
					\begin{tabular}{lll} 
						& $\gamma_\mathrm{T,\,\,sensor\,\,1}$   & $\gamma_\mathrm{T,\,\,sensor\,\,2}$ 
						\tabularnewline
						& (10$^{-7}$ K/V$^2$) & (10$^{-7}$ K/V$^2$)
						\tabularnewline
						\toprule
						trap A ($f_\mathrm{ref}$=15.1\,MHz) &  6.1$\pm$1.6 & 7.2 $\pm$1.4
						\tabularnewline
						trap B ($f_\mathrm{ref}$=24.4\,MHz) & 7.9$\pm$0.9 & 12.1$\pm$0.9
						\tabularnewline
						trap C ($f_\mathrm{ref}$=16.4\,MHz) & 5.2 $\pm$0.2 & 5.9 $\pm$0.2
					\end{tabular}
				\end{ruledtabular}
			\end{center}
		\end{small}
	\end{table}
	
	\noindent In figure~\ref{fig:rf_heating_frequency_dependency_1000V} we show the frequency scaling of the trap temperature rise per kV$^2$ rf voltage amplitude for trap~A (black), trap~B(blue) and trap~C (red). The indicated trend lines show the expected heating according to equation~\eqref{eq:rf_heating_partial} and are in very good agreement with the measurements. The slightly higher heating in trap~A results from the thinner gold coating compared to the other two traps. The lower heating of trap~B with respect to trap~C was discussed previously when comparing the voltage scaling of the two traps.
	The experimental results on the frequency scaling are consistent with the heat distribution obtained from the FEM calculations (section~\ref{thermal analysis}).\\
	The losses in the dielectrics are small and the main contribution of the heating comes from the resistive heating in the thin gold coating as the model and the measured frequency scaling display. Lower heating could be achieved by choosing larger cross-sections for the conductor path or by using thicker layers of deposited gold, at the expense of longer and more difficult laser-machining of the coated chips.\\	
	\noindent The experimental measurements verify the relation in equation~\eqref{eq:rf_heating_partial}, that can be used to estimate the trap temperature rise at any rf drive frequency and amplitude in the tested range of rf frequencies. 
	\section{Trap-related temperature uncertainty budget}
	\label{possible uncertainty budget}
	In this section we discuss the contributions to the uncertainty of the rf-induced temperature rise at the position of the ions and give the resulting absolute uncertainties for the traps B and C.\\
	
	\noindent The uncertainty of the BBR shift stems from the uncertainty of the differential static polarizability of the trapped species, but also from the absolute temperature at the position of the ions, which is defined by the temperature and emissivities of the parts surrounding them. This includes the trap itself, but also the vacuum chamber with its viewports. Even though the stabilization of the latter is a technical challenge, a determination of the BBR temperature at the location of the atoms at the level of 5~mK uncertainty has already been demonstrated in neutral atom experiments~\cite{Ludlow_2015}. The change in thermal radiation due to the temperature rise of the rf feedthrough has also to be taken into account. Its effect on the trap temperature is precisely read-out by the integrated Pt100 sensors but its effect on the thermal radiation should be investigated. We consider in this study only the uncertainty of the trap temperature, which heats up under electrical drive and is often dominating the overall temperature uncertainty in optical ion clocks. The rf-induced temperature rise of the trap with respect to the temperature of the vacuum chamber is obtained from measurements of the Pt100 sensors integrated in the trap and transferred to the position of the ions with equation~\eqref{eq:temperature_ion}.\\
			\begin{figure}[htbp!]
		\includegraphics[width=8.0cm]{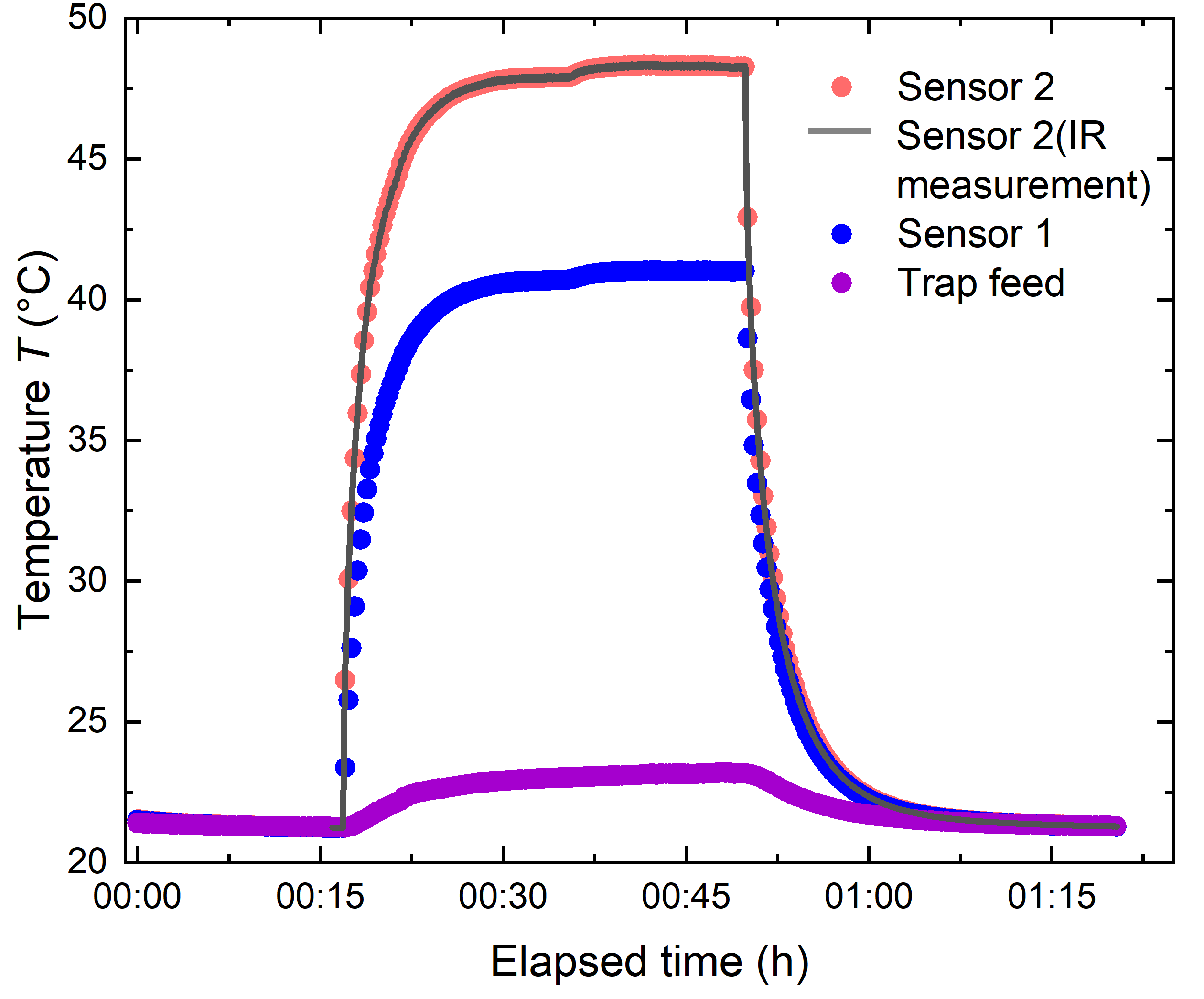}
		\caption{
			Temperature measurement of sensor~1 (blue), sensor~2 (red) and an additional Pt100 sensor placed on one of the trap feet (purple) deduced from their resistances and IR camera measurement (black) at the location of sensor~2. The IR measurement at the location of sensor 2 and its temperature reading agree within the uncertainty of the IR measurement. The heavy heating was due to contamination from the environment, which was cleaned afterwards~\cite{Dolezal_2015}. 
		}
		\label{fig:comparison_Pt100_and_IR_camera}
	\end{figure}

	\noindent The first major uncertainty for the temperature rise at the location of the ions comes from the determination of the temperature rise of the trap and has many contributions. The first are the reading of the integrated Pt100 sensors and the IR camera measurements used as a basis for the numerical model. The Pt100 sensors have been precisely calibrated (see section~\ref{calibration Pt100s}) and an uncertainty $\sigma_\mathrm{T,Pt100}\leq$70\,mK was obtained for typical trap temperatures. Several other sources can contribute to the temperature rise uncertainty. The temperature gradients on a trap chip can lead to additional errors in the temperature estimate. We investigate them via thermal imaging of the trap supplied with 1.5\,kV rf voltage amplitude at 21.5\,MHz (see figure~\ref{fig:IR_measurement_trap_A_1.5kV}). The highest temperature difference on the chip is 0.51\,K. This gradient would lead to high temperature uncertainties, if only a local measurement with the temperature sensors were performed. Fortunately, the numerical model already takes such gradients into account. For the same electric drive, the FEM model gives a maximal temperature gradient of 0.47\,K, which is in good agreement with the measurement. The trap temperature rise given by the model differs slightly from the Pt100 readings. We take this into account with an additional uncertainty on the trap temperature rise. For a conservative approach, we calculate this uncertainty as the maximal local deviation between the numerical model and the experiment. For a voltage amplitude of 1\,kV the maximal deviation between model and experiment is 0.42\,K. This deviation scales quadratically with the rf amplitude so that the corresponding uncertainty is $\sigma_\mathrm{T,FEM/Exp}$=0.42~K/\,kV$^{-2}$. The lower bound of this uncertainty is given by the noise of the IR camera reading, which we recall here from section~\ref{IR measurements}: $\sigma_\mathrm{T,cam}$=0.2\,K+0.05$\Delta T$, where $\Delta T$ is the observed temperature rise in kelvins. The agreement between the Pt100 sensor and IR measurement has already been investigated with the previous trap generation \cite{Dolezal_2015}. There we performed a simultaneous measurement (see figure~\ref{fig:comparison_Pt100_and_IR_camera}) with the integrated Pt100 sensors, an additional Pt100 sensor on one of the trap feet and the IR camera with a region of interest at the location of sensor~2. During the course of the measurement we switched on and off the rf input voltage to see the temperature increase and decrease due to the warm up and cool down of the trap. The IR measurement at the location of sensor~2 and the reading of sensor~2 (black and red line in figure~\ref{fig:comparison_Pt100_and_IR_camera}) agree within the uncertainty of the IR measurement. One should note that the strong heating was due to contamination from the environment, which was cleaned afterwards~\cite{Dolezal_2015}.\\ 

	\noindent In summary the uncertainty on the trap temperature rise $\sigma_\mathrm{\Delta T,Trap_{(i)}}$ measured with sensor i contains the absolute uncertainty of the Pt100 reading and the maximal deviations between the model and experiment $\sigma_\mathrm{T,FEM/Exp}$, leading to the uncertainty $\sigma_\mathrm{\Delta T,Trap_{(i)}}$=$\sqrt{\sigma_\mathrm{T,FEM/Exp}^{2}+\sigma_\mathrm{T,Pt100_{(i)}}^{2}}$ with $\sigma_\mathrm{T,FEM/Exp}\geq $0.25\,K for the typical temperature rises.\\
	
	\noindent Finally, the temperature rise at the location of the sensors $\Delta T_\mathrm{sensor\,\,i}$ is transferred from the Pt100 sensor measurements to the position of the ions $\Delta T_\mathrm{ions}$ via the conversion factors $\alpha$ and $\beta$ derived from the model and introduced in equation~\eqref{eq:temperature_ion}. The uncertainty of those parameters $\sigma_{\alpha}$=$\sigma_{\beta}$=0.10 has been deduced from simulations with varying input parameters (see section~\ref{Temperature derivation at the positions of the ions}).\\
	
	\noindent We use Gaussian error propagation on equation~\eqref{eq:temperature_ion} to determine the final uncertainty of the temperature rise at the position of the ions $\sigma_{\Delta \mathrm{Tions}}$, leading to the expression in equation~\eqref{eq:T_uncertainty_ion}. 
	
	\begin{align}
	\label{eq:T_uncertainty_ion}
	\sigma_{\Delta T\mathrm{ions}}^2= \sigma_{\alpha}^2 \cdot \frac{\Delta T_\mathrm{{sensor\,1}}^2}{4} + \sigma_{\Delta T,\mathrm{Trap_{(1)}}}^2 \cdot \frac{\alpha^2}{4} \nonumber\\
	+ \sigma_{\beta}^2 \cdot \frac{\Delta T_\mathrm{{sensor\,2}}^2}{4} + \sigma_{\Delta T, \mathrm{Trap_{(2)}}}^2 \cdot \frac{\beta^2}{4} 
	\end{align}
	
	\begin{table} [htbp!]
		\caption{Temperature rise at the position of the ions $\Delta T_\mathrm{ion}$ for a rf drive amplitude of 750\,V at typical rf frequencies used in trap~B and~C to trap mixed In$^+$/Yb$^+$ crystals. The values are deduced from the temperature readings of the integrated Pt100 sensors and transferred to the location of the ions with equation~\eqref{eq:temperature_ion}.
		}
		\label{tab:temperature_rise_at_ions}
		\begin{small}
			\begin{center}
				\begin{ruledtabular}
					\begin{tabular}{lll} 
						& Trap~B & Trap~C 
						\tabularnewline
						& $f_\mathrm{rf}$=24.4\,MHz & $f_\mathrm{rf}$=16.4\,MHz
						\tabularnewline
						& $\Delta T_\mathrm{ion}$ (K) &  $\Delta T_\mathrm{ion}$ (K)
						\tabularnewline
						\toprule
						From sensor 1 & 0.19$\pm$0.12 & 0.12$\pm$0.11
						\tabularnewline
						From sensor 2 & 0.22$\pm$0.11 & 0.11$\pm$0.09
						\tabularnewline
						Both sensors (mean) & 0.20$\pm$0.08 & 0.12$\pm$0.07
						\tabularnewline
					\end{tabular}
				\end{ruledtabular}
			\end{center}
		\end{small}
	\end{table}
	\noindent In trap~B and C the rf-induced temperature rise at the position of the ions $\Delta T_\mathrm{ion}$ is maximally 200\,mK at typical trap parameters for trapping Yb$^+$ and In$^+$ ions. Based on the low rf heating of the traps and the good agreement between model and experiment, the achieved uncertainty on the temperature rise at the ions $\sigma_{\Delta \mathrm{Tions}}$ is 80\,mK in trap~B and only 70\,mK in trap~C. This result, which is summarized in table~\ref{tab:temperature_rise_at_ions}, is the smallest reported trap-related temperature uncertainty at the position of the ions and shows the excellent reproducibility of the trap manufacturing. The only trap showing a comparable temperature uncertainty at the position of the ions is the end-cap trap of NPL\cite{Nisbet-Jones_2016} with 140\,mK. That uncertainty is derived from a IR measurement in combination with a FEM model, the trap does not feature an integrated Pt100 sensor for in-situ temperature measurements. Considering the complexity of the design of the scalable chip-based ion traps presented here the achieved uncertainty, derived from the excellent agreement between model and the Pt100 sensor and IR measurements, is even more impressive.
	\begin{table*} [htbp!]
	\caption{Estimate of the trap-related BBR uncertainty of trap~B for different ion species. The rf voltage amplitude has been calculated to fit published secular frequencies (for the clock ion species) used in current optical clocks. The corresponding average temperature rise $\Delta T_{\mathrm{{sensor\,1\,+\,2}}}$ on sensor 1 and 2 and at the position of the ions $\Delta T_{\mathrm{ions}}$ is an estimate obtained by rescaling the trap heating for rf amplitude and frequency as described in section~\ref{Pt100s measurements}.}	
	\label{tab:comparison_BBR_different_clock_species}
	\begin{small}
		\begin{center}
			\begin{ruledtabular}
				\begin{tabular}{lllllll} 
					Species & rf frequency & rf amplitude & Radial secular & $\Delta T_{\mathrm{\,sensor\,1\,+\,2}}$ & $\Delta T_{\mathrm{ions}}$ & Trap-related
					\tabularnewline
					(clock/logic ion,  & (MHz) & (V) & frequency  & (K) & (K) &  BBR uncertainty
					\tabularnewline
					transition) &  &  & (MHz) & & & $\sigma_\Delta \nu/\nu_0 $ (10$^{-20}$) at 22$^{\circ}$C
					\tabularnewline
					\toprule
					$^{115}$In$^+$ ($^{172}$Yb$^+$) \cite{Keller_2019_PRA} & 16.4 & 524 & 0.75 & 0.12$^{+0.25}_{-0.12}$ & 0.05$^{+0.07}_{-0.05}$ & 1.2
					\tabularnewline
					$^{115}$In$^+$ ($^{172}$Yb$^+$) \cite{Keller_2019_PRA} & 16.4 & 1060 & 1.5 & 0.51$\pm$0.48 & 0.19$\pm$0.13 & 2.3
					\tabularnewline
					$^{27}$Al$^+$ ($^{25}$Mg$^+$) \cite{Brewer_2019} & 40.72 & 1650 & 4.0 & 7.86$\pm$1.15 & 2.85$\pm$0.65 & 3.2
					\tabularnewline
					$^{27}$Al$^+$ ($^{25}$Mg$^+$) \cite{Xu_2016} & 23.9 & 481 & 2.0 & 0.22$^{+0.25}_{-0.22}$ & 0.08$\pm$0.07 & 3.6
					\tabularnewline
					$^{27}$Al$^+$  ($^{40}$Ca$^+$) \cite{Hannig_2019} & 24.65 & 936 & 3.75 & 0.89$\pm$0.37 & 0.32$\pm$0.12 & 0.57
					\tabularnewline
					$^{27}$Al$^+$  ($^{40}$Ca$^+$) \cite{Shang_2016} & 17.128 & 520 & 3.0 & 0.13$^{+0.25}_{-0.13}$ & 0.05$^{+0.07}_{-0.05}$ & 0.33
					\tabularnewline
					$^{88}$Sr$^+$ \cite{Dube_2015} & 14.408 & 568 & 1.2 & 0.11$^{+0.25}_{-0.11}$ & 0.04$^{+0.07}_{-0.04}$ & 48
					\tabularnewline
					$^{176}$Lu$^+$~($^{1}S_{0}-^{3}D_{1}$) \cite{Tan_2019} & 16.74 & 767 & 0.7 & 0.28$\pm$0.25 & 0.10$\pm$0.07 & 0.13
					\tabularnewline
					$^{176}$Lu$^+$~($^{1}S_{0}-^{1}D_{2}$) \cite{Kaewuam_2019} & 16.8 & 640 & 0.585 & 0.20$\pm$0.25 & 0.07$\pm$0.07 & 1.7
					\tabularnewline
					$^{171}$Yb$^+$ \cite{Tamm_2009} (E3) & 15 & 668 & 0.7  & 0.17$^{+0.25}_{-0.17}$ & 0.06$^{+0.07}_{-0.06}$ & 6.2
					\tabularnewline
					$^{171}$Yb$^+$ \cite{Tamm_2009} (E2) & 15 & 944 & 0.7  & 0.17$^{+0.25}_{-0.17}$ & 0.06$^{+0.07}_{-0.06}$ & 51
					\tabularnewline
					$^{171}$Yb$^+$ \cite{Nisbet-Jones_2016} (E3) & 14 & 895 & 1.0  & 0.27$^{+0.34}_{-0.27}$ & 0.10$\pm$0.09 & 8.6
					\tabularnewline
					$^{40}$Ca$^+$ \cite{Cao_2017}& 24.54 & 771 &  2.1 & 0.60$\pm$0.25 & 0.22$\pm$0.08 & 95.2
					\tabularnewline
				\end{tabular}
			\end{ruledtabular}
		\end{center}
	\end{small}
\end{table*}
\section{BBR shift in typical ion clock species}
\label{BBR_shift_ion_species}
To benchmark our trap, we give an estimate on the trap-related BBR shift uncertainty for candidates for optical ion clocks. For attaining appropriate comparisons, we use published values for the rf and secular frequencies and adapt the rf voltage amplitude to obtain the same confinement in our trap. Based on the ability to calculate the rf-induced heating depending on frequency and amplitude with equation~\eqref{eq:rf_heating_partial}, we calculate the related trap temperature rise as an example for AlN trap~B at the location of the two sensors $\Delta T_{\mathrm{\,sensor\,1\,+\,2}}$ and at the position of the ions $\Delta T_{\mathrm{ions}}$ with equation~\eqref{eq:temperature_ion}. Table~\ref{tab:comparison_BBR_different_clock_species} summarizes this estimate.\\

\noindent For rf drive frequencies below 25\,MHz and amplitudes below 1.5\,kV, the temperature rise at the position of the ions in trap~B is below 700\,mK and its uncertainty on the order of a few 100\,mK maximum. The uncertainty is given by the calibration of the integrated Pt100 sensors, the deviation between the thermal numerical model and experiment and the conversion factors $\alpha$ and $\beta$ relating the temperature measured with the two sensors to the position of the ions, as explained in section~\ref{possible uncertainty budget}. For the clock candidates with extremely low differential polarizabilities, like $^{27}$Al$^+$, $^{115}$In$^+$ and $^{176}$Lu$^+$, the resulting trap-related BBR-shift uncertainty in our trap~B is in the 10$^{-20}$ range or below. Even for ion species with extremely high polarizability the resulting trap-related BBR-shift uncertainty is below 2$\times$10$^{-18}$. Trap~B and C are identical, but are mounted differently in the vacuum chamber. The resulting trap-related BBR uncertainties in trap~C correspond to those of trap~B within 10\% deviation. 
\section{Summary and outlook}
\label{conclusion}
We presented scalable chip ion traps with very low and precisely determined heating in operation, thus obtaining one of the smallest trap-related contribution to the BBR uncertainty of trapped ions experiments. We described design considerations for the development of ions traps with optimized passive thermal management and discussed possible materials for trap fabrication. The traps are composed of a stack of four AlN chips and feature two integrated Pt100 sensors for a precise determination of the trap temperature.
We used FEM calculations to model the thermal heating of the trap and analyze the heat distribution between its parts. The model is precisely refined using IR measurements of the related material emissivities and imaging of the trap under use. Controlled local heating is applied on the trap to derive the thermal contacts that are not accessible via IR imaging only. Conversion factors are derived from the model in order to determine precisely the rf-induced temperature rise seen by the ions from the reading of the integrated Pt100 sensors. The factors are used to obtain a real-time in-situ estimate of the  temperature rise at the position of the ions.\\

\noindent In total, three traps of the same geometry have been analyzed experimentally. We described the calibration of the sensors of two traps to absolute uncertainties around 70\,mK at a typical lab temperature around 22\,$\degree$C.
The temperature rise of all three traps has been investigated using the integrated Pt100 sensors and for one trap IR measurements have been done in parallel to compare the results and to analyze the maximal temperature gradient across the trap. We investigated the scaling of the heating with rf frequency and amplitude and found excellent agreement with the numerical model. Hereby we verify that the model can be used to estimate the expected trap temperature rise at any rf frequency or amplitude. All three analyzed traps display very low rf-induced heating. Typical trap parameters to trap mixed In$^+$/Yb$^+$ crystals lead to a temperature rise at the ions position of 200\,mK in trap~B and C as shown in table~\ref{tab:temperature_rise_at_ions}.\\

\noindent Finally, we discussed the contributions to the uncertainty of the temperature at the position of the trapped ions. The temperature at the position of the ion is dominated by the contribution from the trap: The reasons are that a large part of the ions solid angle is covered by the trap and the distance between trap and ion is on the order of a tenth of a millimeter to a few millimeter. The temperature uncertainty at the position of the ions related to our trap is below 80\,mK at typical trap parameters to confine mixed In$^+$/Yb$^+$ crystals, when using both sensors as summarized in table~\ref{tab:temperature_rise_at_ions}.
This corresponds to one of the smallest published trap-related temperature uncertainties among the trapped ion community and is an important achievement for complex, scalable chip ion traps. 
Together with the suitability of the traps for high-precision spectroscopy of several ions\cite{Keller_2019_PRA}, the low trap-related temperature uncertainty paves the way for multi-ion or quantum logic clocks with extremely low BBR-shift uncertainties. The estimate of the trap-related BBR-shift uncertainty for typical ion clock species in our trap is in the 10$^{-19}$ range or below for clock transitions with low differential static polarizability. This demonstrates that even with the use of complex chip-based ion traps the overall uncertainty of the best modern single-ion clocks is not limited by the trap-related BBR-shift uncertainty. To give the full uncertainty of the temperature at the ion's position, one also has to consider the temperature of the vacuum chamber. For individual setups the specific vacuum chamber and rf feedthrough can be included in the model. A control of the uncertainty of the temperature of the vacuum chamber on a level of 5\,mK has already been achieved.\cite{Ludlow_2015}\\

\noindent For many ion clock species the uncertainties on the differential static polarizability $\Delta \alpha_\mathrm{stat}$ and the dynamical correction $\eta$ are limiting the total BBR-shift uncertainty since theoretical values are difficult to derive with low uncertainty\cite{Safranova_2011} and experimental characterization is sometimes challenging. The unknown polarizability is a limiting factor for absolute frequency measurements, and its precise determination is required for a possible redefinition of the second.\cite{Riehle_2015} Many applications, however require just relative frequency measurements. One example is the use of optical clocks to determine geodetic height differences, referred to as chronometric leveling.\cite{Mehlstaeubler_2018, Bjerhammar_1975, Bjerhammar_1985} Frequency comparisons using portable clocks to evaluate the time differences at different geodetic heights require only reproducible clock measurements.\cite{Mura_2013, Ohtsubo_2019, Cao_2017, Grotti_2018, Koller_2017, Bongs_2015, Takamoto_2020, Kong_2020, Hannig_2019} Also the relative frequency comparison of two clocks using the same ion species located at different geodetic heights directly measures the height difference without prior knowledge of the polarizability. The here-demonstrated thermal management of ion traps enables a relative frequency comparison with a reproducibility in the 10$^{-19}$ range with respect to the BBR shift. For some species even the 10$^{-20}$ range is possible (according to table~\ref{tab:comparison_BBR_different_clock_species}). The resolution of a frequency difference in the 10$^{-19}$ range corresponds to a physical height resolution in the millimeter regime.\\

\noindent The dominating uncertainty contribution for the trap-related temperature uncertainty is the conservative maximal deviation between the model and the experimental results. Additional IR measurements with more temperature references, together with a more detailed fitting of the thermal contacts inside the trap stack could drastically reduce this uncertainty. Alternatively, the use of crystalline materials, such as diamond and sapphire for the fabrication of the trap can provide even better precision owing to their higher homogeneity.\\
\begin{acknowledgments}
We thank Physikalisch-Technische Bundesanstalt department 5.5 for collaboration on trap fabrication, and S.A. King and H. Liu for helpful comments on the manuscript. This work originated from a collaborative project within the European Metrology Research Programme (EMRP) SIB04. We acknowledge funding by the Deutsche Forschungsgemeinschaft (DFG) through grant CRC SFB 1227 (DQ-mat, project B03), through Germany’s Excellence Strategy EXC2123 QuantumFrontiers and by the BMBF under grant 13N14962. The CMI participation in this project was funded by Institutional Subsidy for Long-Term Conceptual Development of a Research Organization granted to the CMI by the Ministry of Industry and Trade.
\end{acknowledgments}

\section*{Data availability}
The data that support the findings of this study are available from the corresponding author upon reasonable request.
	
\section*{Appendix: Calibration of the integrated Pt100 sensors}
\label{appendix_calibration}
The uncertainty of the Pt100 sensors is specified by the manufacturer to be 300\,mK. For a more precise temperature estimate a smaller uncertainty is necessary. The calibration of bare sensors is commonly carried out with an oil-bath. When calibrating the sensors integrated into the trap, it is necessary to include the additional conductive paths on the trap and the electrical contacts between sensors and trap. The calibration must therefore be done with the complete trap. To maintain the UHV suitability we developed a complex calibration procedure avoiding the use of the common oil-bath for this type of calibration. It is described in the following. The sensors make use of the temperature dependent electrical resistance $R(T)$ which can be approximated with a second-order polynomial (T in $\degree$C):
\begin{align}
\label{eq:resistance_Pt100}
R(T)=R_0 \cdot (1+c_1 \cdot T+c_2 \cdot T^2). 
\end{align}
For temperatures above 0\,$\degree$ C the inverse relation is given by:
\begin{align}
\label{eq:temperature_Pt100}
T(R)=\frac{-R_0 \cdot c_1\sqrt{(R_0 \cdot c_1)^2-4R_0 \cdot c_2 \cdot (R_0-R)}}{2R_0 \cdot c_2}.
\end{align}
To achieve a low uncertainty for the temperature measurement, the coefficients $c_1$, $c_2$ and $R_0$ have been determined for the sensors on trap~B and C.\\

\begin{figure}[htbp!]
	\includegraphics[width=8.5cm]{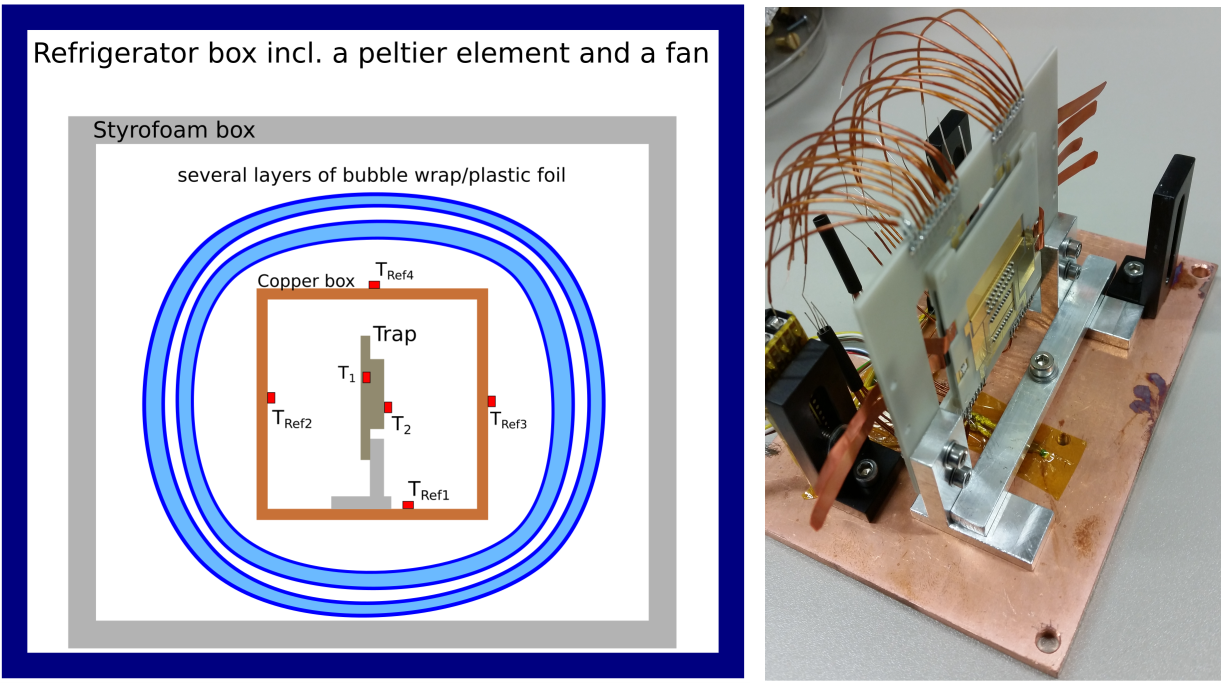}
	\caption{Set-up for the absolute temperature calibration of the integrated Pt100 sensors. \textit{Left:} side view of the trap placed inside a refrigerator box including a Peltier module. \textit{Right:} Picture of the trap placed on the copper base plate. The rest of the box is then placed on top of it.
	}
	\label{fig:Pt100_calibration_setup}
\end{figure}

\noindent \textbf{Calibration procedure}\\
For this calibration the trap is installed inside a copper cuboid with a wall thickness of 10\,mm. Four reference sensors, calibrated at PTB to an uncertainty of 7\,mK\footnote{Cleverlab}, are placed on the inside and outside of the cuboid. The cuboid is wrapped in several layers of bubble wrap and is placed inside a Styrofoam box. This box is positioned inside a refrigerator box with an integrated Peltier element, which is used to heat and cool the box to different temperatures (figure~\ref{fig:Pt100_calibration_setup}). The resistance measurements of the six sensors, the two on the trap and the four reference sensors, is performed with a calibrated precision digital multimeter\cite{agilent} and the four-wire sensing method. In total the trap in its thermal shield has been cooled and heated to eight different temperatures between 19.7\,$\degree$C and 34\,$\degree$C. 

\noindent Before measuring the resistances of the six sensors for the respective temperatures the system was allowed to equilibrate for two days. The resistance value of each sensor at each temperature is the mean of 25 equidistant measurements over a time interval of two minutes while its standard deviation gives a statistical uncertainty for that value. The resistance measurement is performed twice for one of the reference sensors, at the beginning and at the end of each measurement series, in order to quantify the temperature drift over the time elapsed during the measurement procedure for all sensors. The mean value of the four calibrated reference sensors is taken as the equilibrated temperature inside the box. The reading of the Pt100 sensors is done with a current of 1\,mA leading to a resistive heating, which needs to be considered for precise temperature measurements. We characterize this so called self-heating for all sensors and consider it for the temperature and uncertainty determination. We apply small dc currents between 0.5\,mA and 10\,mA to the sensors and measure the current and voltage drop across the sensor with two calibrated precision digital multimeters \cite{agilent}. With Ohm's law the respective resistance and thereby the temperature can be calculated and the quadratic behavior between measurement current and temperature can be fitted.  \\

\begin{figure}[htbp!]
	\includegraphics[width=8.5cm]{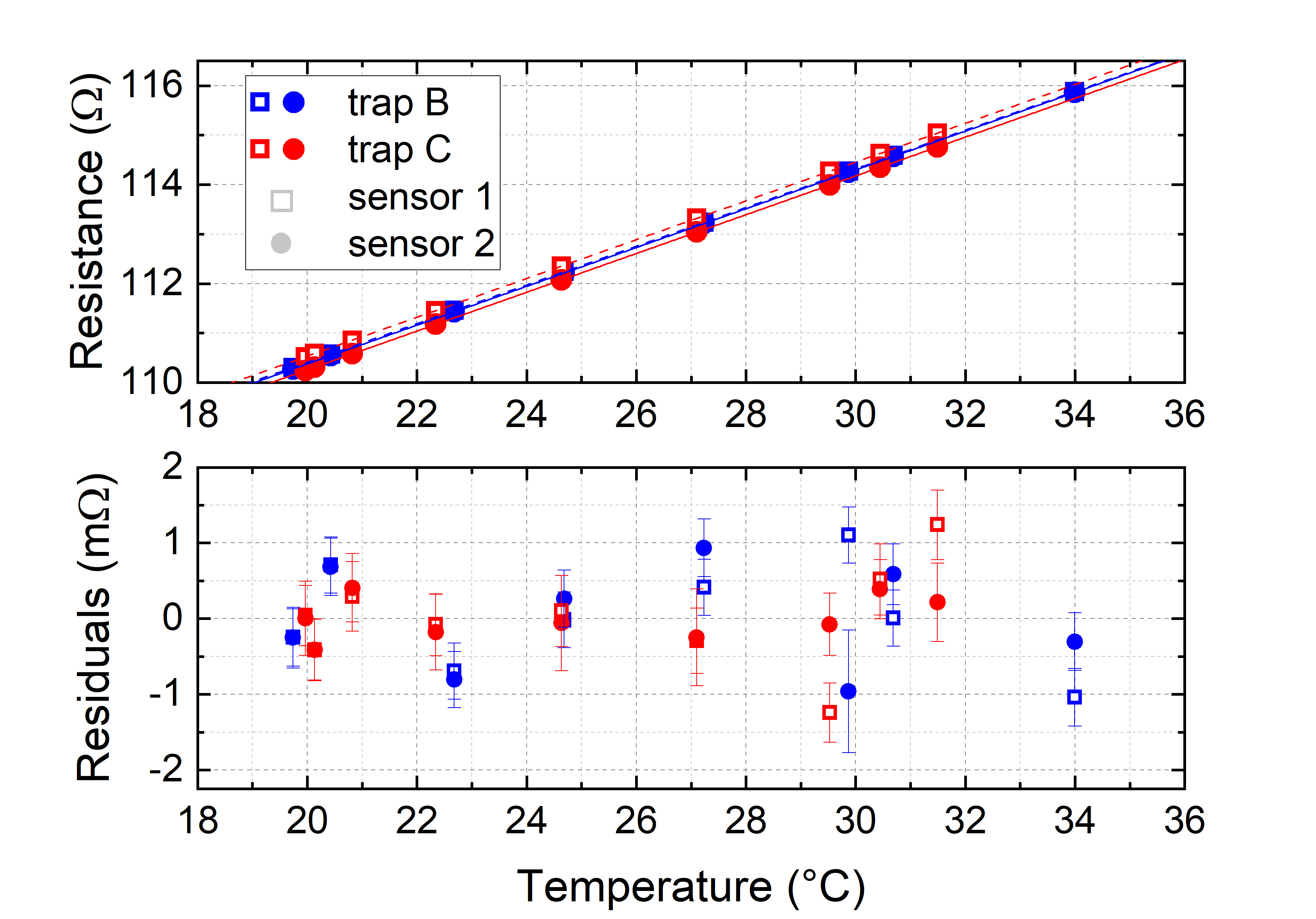}
	\caption{\textit{Top:} Temperature dependency of the resistance of the integrated Pt100 sensors on trap~B (blue) and C (red). The open symbols correspond to sensor~1 and the closed to sensor~2. The error bars are below 0.5\,m$\Omega$ and thereby smaller than the plotted points. \textit{Bottom:} Residuals of the fitted measurement. The temperature uncertainties across the temperature range of 20-34\,$\degree$C are between 9-26\,mK for trap~B and 4-31\,mK for trap~C.
	}
	\label{fig:Pt100_calibration_trap_B_and_C}
\end{figure}

\noindent \textbf{Result}\\
Figure~\ref{fig:Pt100_calibration_trap_B_and_C} shows the result of the calibration of the integrated sensors for the traps B and C and the fit of the data using equation~\eqref{eq:resistance_Pt100}. Trap A has yet to be calibrated. Table~\ref{tab:Pt100_calibration} lists the respective fit results of the parameters $c_1$, $c_2$ and $R_0$.

\begin{table} [htbp!]
	\caption{Calibration coefficients for the integrated PtT100 sensors of trap~B and C.
	}
	\label{tab:Pt100_calibration}
	\begin{small}
		\begin{center}
			\begin{ruledtabular}
				\begin{tabular}{llll} 
					Trap & $R_0$ ($\Omega$) & $c_1$ (10$^{-3}$\,$\Omega$/$\degree$C) & $c_2$ (10$^{-7}$\,$\Omega$/$\degree$C)
					\tabularnewline
					\toprule
					B - Sensor 1& 102.530$\pm$0.011 &3.860$\pm$0.009  &-8.9$\pm$1.7
					\tabularnewline
					B - Sensor 2 & 102.478$\pm$0.012 & 3.872$\pm$0.010 & -8.8$\pm$1.8 
					\tabularnewline
					C - Sensor 1 & 102.658$\pm$0.0101& 3.850$\pm$0.008 & -6.1$\pm$1.6 
					\tabularnewline
					C - Sensor 2 & 102.376$\pm$0.010& 3.864$\pm$0.008 &-7.1$\pm$1.6
					\tabularnewline
				\end{tabular}
			\end{ruledtabular}
		\end{center}
	\end{small}
\end{table}

\noindent Table~\ref{tab:Pt100_uncertainty_budget} lists all accounted uncertainties for the temperature determinations that are used for the calibration of the two integrated Pt100 sensors. The dominant uncertainty source at 22\,$\degree$C (as a typical lab temperature) are the temperature gradients across the copper cuboid which increase with the temperature at a rate of 3\,mK/$\degree$C. 

\begin{table} [htbp!]
	\caption{Uncertainty contributions for the temperature determination at T=22\,$\degree$C within the Pt100 sensor calibration. The resulting absolute uncertainty of the temperature determination $\sigma_\mathrm{T, Pt100}$ is obtained with Gaussian error propagation on equation~\eqref{eq:temperature_Pt100}
	}
	\label{tab:Pt100_uncertainty_budget}
	\begin{small}
		\begin{center}
			\begin{ruledtabular}
				\begin{tabular}{ll} 
				Source& Uncertainty (mK)
			\tabularnewline
			\toprule
			Reference Pt100 sensors & $<$8
			\tabularnewline
			Digital multimeter & $<$1
			\tabularnewline
			Self-heating per sensor & $<$1
			\tabularnewline
			Temperature drift during measurements & $<$4
			\tabularnewline
			Temperature gradients over copper box & $<$10
			\tabularnewline
				\end{tabular}
			\end{ruledtabular}
		\end{center}
	\end{small}
\end{table}
\noindent Less dominant are the temperature uncertainties of the individual reference sensors and the temperature drift during the measurement. The uncertainties of the digital multimeter and the self-heating of all sensors are almost negligible. Table~\ref{tab:Pt100_uncertainty_budget} lists only the uncertainty of a resistance measurement at a single temperature. The temperature dependency of the sensor resistance is then fitted using equation~\eqref{eq:resistance_Pt100}. The fit errors on the parameters $c_1$, $c_2$ and $R_0$ add to the total temperature uncertainty of the sensors.\\

\noindent With Gaussian error propagation on equation~\eqref{eq:temperature_Pt100} and the calibration coefficients from table~\ref{tab:Pt100_calibration} we derive the absolute temperature uncertainty $\sigma_\mathrm{T, Pt100}$. The resulting absolute temperature uncertainty $\sigma_\mathrm{T, Pt100}$ for both integrated sensors in trap~B and C is shown in figure~\ref{fig:absolute_temperature_uncertainty} in section~\ref{calibration Pt100s}. For a typical lab temperature of 22\,$\degree$C the absolute temperature uncertainty of both sensors is below 70\,mK in the case of trap~B and below 60\,mK for trap~C.

\end{document}